\documentclass[journal]{IEEEtran}
\usepackage{amsmath,amsfonts}
\usepackage{algorithmic}
\usepackage{algorithm}
\usepackage{array}
\usepackage{caption}
\usepackage{subcaption}
\usepackage{textcomp}
\usepackage{stfloats}
\usepackage{url}
\usepackage{verbatim}
\usepackage{graphicx}
\usepackage{cite}
\usepackage{amssymb}
\usepackage{mathtools}
\usepackage{amsmath}
\usepackage{enumitem}
\usepackage{multirow}

\def\etal{\emph{~et~al. }}
\makeatother

\ifCLASSINFOpdf
\else
\fi

\begin{document}



\title{AROW: V2X-based Automated Right-of-Way Algorithm for Cooperative Intersection Management}


\author{Ghayoor Shah$^{1}$, Danyang Tian$^{2}$, Ehsan Moradi-Pari$^{2}$, Yaser P. Fallah$^{1}$%
\thanks{$^{1}$ Connected \& Autonomous Vehicle Research Lab (CAVREL), University of Central Florida, Orlando, FL, USA. \tt\small {ghayoor.shah@ucf.edu}}%
\thanks{$^{2}$ Honda Research Institute, Ann Arbor, MI}%
}%



\maketitle

\begin{abstract}
Research in Cooperative Intersection Management (CIM), utilizing Vehicle-to-Everything (V2X) communication among Connected and/or Autonomous Vehicles (CAVs), is crucial for enhancing intersection safety and driving experience.
CAVs can transceive basic and/or advanced safety information, thereby improving situational awareness at intersections.
The focus of this study is on unsignalized intersections, particularly Stop Controlled-Intersections (SC-Is), where one of the main reasons involving crashes is the ambiguity among CAVs in SC-I crossing priority upon arriving at similar time intervals.
Numerous studies have been performed on CIM for unsignalized intersections based on centralized and distributed systems in the presence and absence of Road-Side Unit (RSU), respectively. 
However, most of these studies are focused towards replacing SC-I 
where the scheduler provides
spatio-temporal or sequence-based reservation to CAVs, or
where 
it
controls CAVs via kinematic commands.
These methods cause CAVs to arrive at the intersection at non-conflicting times and cross without stopping. 
This logic is severely limited in real-world mixed traffic comprising human drivers where kinematic commands and other reservations cannot be implemented as intended.
Thus, given the existence of SC-Is and mixed traffic, 
it is significant to develop CIM systems 
incorporating
SC-I rules while assigning crossing priorities and resolving the related ambiguity.
In this regard, we propose a distributed Automated Right-of-Way (AROW) algorithm for CIM 
to assign explicit SC-I crossing turns to CAVs and mitigate hazardous scenarios due to 
ambiguity towards crossing priority.
The algorithm is validated with extensive experiments for its functionality, scalability, and robustness towards CAV non-compliance, and it outperforms the current solutions.

\end{abstract}

\begin{IEEEkeywords}
Intersection Management, 
Connected Vehicle Applications,
V2X, V2V, Right-of-Way, Stop-Controlled Intersections
\end{IEEEkeywords}

\section{Introduction}
\label{Section1}
\IEEEPARstart{I}{ntersection} management is one of the most significant and challenging topics for safe and efficient driving experience. 
Intersections can be mainly divided into two categories, i.e., signalized and unsignalized intersections. Signalized intersections comprise of traffic lights, whereas the unsignalized intersections consists of stop or yield signs, or in some instances, no signs at all.
The ever-increasing vehicles on roads can lead to congestion~\cite{rafter2017traffic}, traffic delays~\cite{rahmati2017towards}, and economic and societal costs due to crashes~\cite{blincoe2015economic}.
In this context, unsignalized~\cite{retting2003analysis} and signalized~\cite{shi2016intelligent} intersections have been found to be hazardous and inefficient towards catering large volumes of traffic. 

According to the United States National Motor Vehicle Causation Survey, around 35\% of the crashes between 2005 and 2007 occurred on intersections, out of which, insufficient situational awareness and ambiguity over intersection priority constituted for 44.1\% and 8.4\%, respectively~\cite{choi2010crash}. 
Similarly, it was reported that in the United States and Europe, around 40\% of all traffic accidents occurred at intersections~\cite{azimi2014stip}.
Based on the National Highway Traffic Safety Administration (NHTSA) report~\cite{singh2015critical} conducted in 2015, around 95\% of the total vehicle crashes in the US from 2005 to 2007 were a consequence of human error, out of which, around 35\% occurred due to bad driver decisions. 
Given these studies, it is critical to investigate intelligent traffic management systems. 

In this regard, Cooperative Intersection Management (CIM) systems have been a major source of interest in the research 
community~\cite{chen2015cooperative, namazi2019intelligent, zhong2020autonomous,gholamhosseinian2022comprehensive}
for signalized and unsignalized intersections.
CIM systems are used by Connected and/or Autonomous Vehicles (CAVs) to improve the overall driver experience.
CAVs employ Vehicle-to-Everything (V2X) communication~\cite{jkenney:dsrcmain, 3gpp:3gppmain, shah2019real, shah2020rve} using their On-Board Units (OBUs) to enhance situational awareness and improve safety standards while making driving decisions. 
Although there is abundant literature on CIM for signalized intersections~\cite{schmocker2008multi,devoe2008distributed,chang2013study,buinevich2021v2x,  gutesa2021development,parks2022intersection}, the focus of this study is on 
unsignalized intersections, particularly Stop Controlled-Intersections (SC-Is) which are major source of crashes in unsignalized intersections~\cite{retting2003analysis}. One of the main reasons involving SC-I crashes is the ambiguity in SC-I crossing priority~\cite{choi2010crash} when multiple CAVs approach the SC-I at similar time intervals. 
To the authors' best knowledge and as explained below, none of the related works focus on unsignalized CIM systems explicitly for SC-I with a focus on ambiguity mitigation, which is one of the main objectives of this work.

CIM systems for unsignalized intersections can be generally divided into two categories: centralized and distributed CIM.
Centralized CIM systems utilize Road-Side Unit (RSU) to function as an intersection manager for passing priorities and information about nearby CAVs.
On the other hand, distributed CIM systems solely rely on inter-vehicular communication and choose one of the CAVs as an arbitrator that can assign intersection priorities without the usage of any RSU.
Numerous studies exist for centralized unsignalized CIM 
~\cite{dresner2004multiagent, kowshik2011provable,perronnet2012cooperative, wu2012cooperative,li2020intersection,gregoire2016hybrid}.
However, the main issue with centralized CIM is that RSUs can suffer deadlocks in terms of memory limitations in the case of high density traffic, thereby making them unfeasible in real-life scenarios. Secondly, RSU equipment needs to be installed on all unsignalized intersections 
in addition to the OBUs within CAVs, hence these systems are costly and currently offer limited scalability on roads.
Given these issues, distributed CIM~\cite{vanmiddlesworth2008replacing, li2006cooperative,katriniok2019nonlinear, hassan2014fully} offers an economic and scalable alternative 
since it does not rely on RSUs for its operation, and hence is the basis of this work as well.
One common feature in most existing centralized and distributed algorithms for unsignalized intersections is that they are aimed at replacing the SC-I. 
This is achieved by the scheduler either providing spatio-temporal or sequence-based reservation of intersection, or by requiring control of CAVs.
Requiring control refers to the assigned scheduler sending
kinematic commands instructing
CAVs to accelerate or decelerate varyingly so as to arrive at the intersection at non-conflicting times.
By using any of these techniques, the main aim is to mitigate any collision scenarios and allow the CAVs to possibly pass the intersection without adhering to the general SC-I right-of-way rules of stopping before passing~\cite{mutcd}.
Although these solutions are effective for fully autonomous traffic, however, they are severely limited in 
mixed traffic comprising of autonomous and human-operated vehicles, where kinematic commands or spatio-temporal reservation cannot be strictly implementable by human drivers.
Given that SC-Is continue to be an integral part of traffic control and can encompass mixed traffic, it is significant to develop CIM systems that incorporate the rules of SC-I and are focused towards assigning passing priorities 
with the aim of mitigating ambiguity among CAVs when they arrive at the SC-I at similar times.

\begin{table}[t]
\setlength{\tabcolsep}{1pt} 
  \centering
  \caption{\small{Tabular comparison of characteristics of AROW and related works. 'hybrid' refers to network containing both centralized and distributed system.}}
  \resizebox{0.49\textwidth}{!}{
      \begin{tabular}{|c|c|c|c|c|}
        \hline
         \multicolumn{1}{|c|}{
         \textbf{Works}}
         & 
         \multicolumn{1}{c|}{\textbf{Intersection}} & \multicolumn{1}{c|}{\textbf{Network}} & 
         \multicolumn{1}{c|}{\textbf{SC-I based?}} &
         \multicolumn{1}{c|}{\textbf{Considers mixed autonomy CAVs?}} 
         \\
         \hline
         ~\cite{schmocker2008multi} & signalized & centralized & 
        N/A & $\times$
        \\
        \hline
        ~\cite{devoe2008distributed} & signalized & centralized & 
        N/A & $\times$
        \\
        \hline
         ~\cite{chang2013study} & signalized & hybrid & 
         N/A & \checkmark
         \\
        \hline
        ~\cite{buinevich2021v2x} & signalized & hybrid & 
         N/A & \checkmark 
         \\
        \hline
        ~\cite{gutesa2021development} & signalized & centralized & 
         N/A & \checkmark 
         \\
        \hline
        ~\cite{parks2022intersection} & signalized & centralized & 
         N/A & \checkmark 
         \\
        \hline
        ~\cite{dresner2004multiagent} & unsignalized & centralized & 
        $\times$ & $\times$
        \\
        \hline
        ~\cite{kowshik2011provable} & unsignalized & hybrid & 
        $\times$ & $\times$
        \\
        \hline
        ~\cite{perronnet2012cooperative} & unsignalized & centralized & 
        $\times$ & \checkmark
        \\
        \hline
        ~\cite{wu2012cooperative} & unsignalized & centralized & 
        $\times$ & \checkmark
        \\
        \hline
        ~\cite{li2020intersection} & unsignalized & centralized & 
        $\times$ & $\times$
        \\
        \hline
        ~\cite{gregoire2016hybrid} & unsignalized & hybrid & 
        $\times$ & $\times$
        \\
        \hline
        ~\cite{vanmiddlesworth2008replacing} & unsignalized & distributed & 
        $\times$ & $\times$
        \\
        \hline
        ~\cite{li2006cooperative} & unsignalized & distributed & 
        $\times$ & $\times$
        \\
        \hline
        ~\cite{katriniok2019nonlinear} & unsignalized & distributed & 
        $\times$ & $\times$
        \\
        \hline
        ~\cite{hassan2014fully} & unsignalized & distributed & 
        $\times$ & $\times$
        \\
        \hline
        \textbf{AROW} & unsignalized & distributed & 
        \checkmark & \checkmark
        \\
       \hline
      \end{tabular}
  }
  \label{table:relatedworks_3.2}
\end{table}

Therefore, to address this, we propose a distributed CIM algorithm 
for SC-Is
namely Automated Right-of-Way (AROW) that utilizes an earlier designed and implemented Driver Messenger System 
(DMS)~\cite{shah2022enabling, tian2023systems} 
based on the relevant source material~\cite{rajab2017driver}.
CAVs can employ AROW in SC-I to negotiate SC-I crossing priorities in the form of explicit passing turns through Driver Intention Messages (DIMs) and avoid any potential for ambiguity. The proposed algorithm is robust to any non-compliance from CAVs and not only assists in mitigating ambiguity, but also in promoting fairness when it comes to SC-I crossing, even for large-scale CAV scenarios.

Our contributions are as follows:
\begin{itemize}




\item We propose a V2X-based distributed algorithm AROW for CIM at SC-Is to mitigate hazardous traffic scenarios arising due to ambiguity towards SC-I crossing priority,
which is achieved by transmitting a series of DIMs where CAVs arbitrate and assign distinct SC-I passing priorities to each other.
The distributed nature of AROW allows it to be scalable for high density traffic scenarios which may not be possible using a centralized RSU given its communication bandwidth and memory limitations.


\item AROW is the first distributed algorithm for CIM that does not require low-level vehicle control such as specific kinematic commands, and 
is equipped to abide by the SC-I rules and allow explicit passing turns, while handling any unpredictability caused by driver behavior.
This allows AROW to be feasible for mixed traffic scenarios as opposed to other works in the literature.

\item The proposed AROW algorithm is capable of handling any level of non-compliance by CAVs in terms of SC-I crossing priorities and outperforms related methods from the literature in terms of ambiguity mitigation.

\end{itemize}

\begin{figure}[t]
\centerline{\includegraphics[trim=5 0 5 0,clip,width=.495\textwidth]{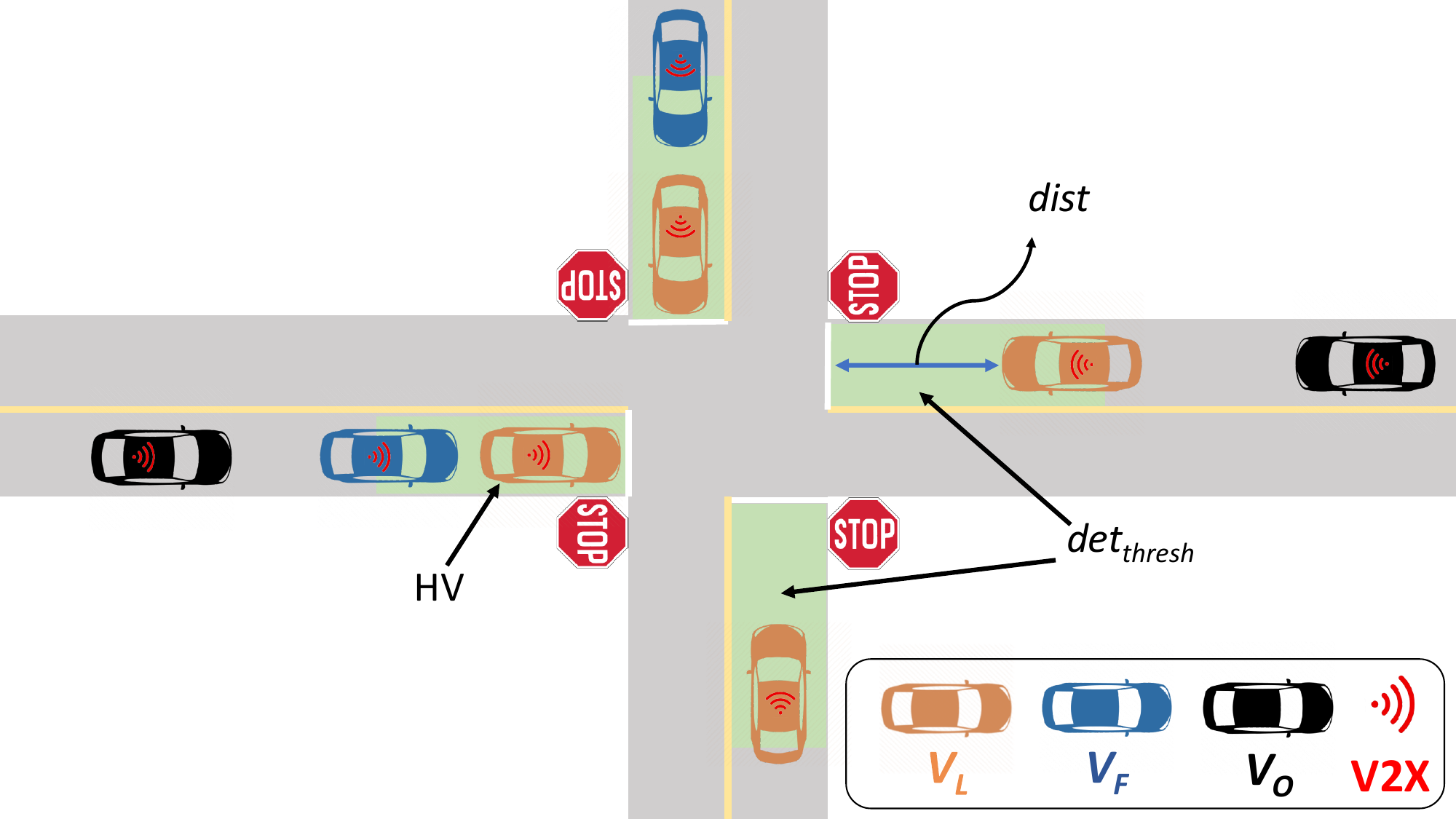}}
\caption{\small{Bird-eye view of a Stop Controlled-Intersection (SC-I) during application detection phase.}}
\label{fig:sci_det}
\end{figure}

\section{Related Works}
CIM systems for unsignalized intersections can be divided into centralized and distributed systems. 
Although there are
numerous studies in the literature 
as mentioned below
where
CIM systems provide safe and efficient solutions for unsignalized intersections in general, none are specifically focused towards SC-Is and the inherent ambiguity in crossing priority,
specially in mixed traffic scenarios.
This is precisely the focus of this study by proposing AROW algorithm that can be operated within the DMS framework.
Table \ref{table:relatedworks_3.2} compares the characteristics of AROW with the related works.

\subsection{Centralized CIM for unsignalized intersections}
Numerous studies offer different centralized CIM algorithms.
Dresner~\etal~\cite{dresner2004multiagent} utilize cooperative resource reservation where CAVs reserve space tiles in the vicinity of the intersection at specific times, thereby ensuring no conflict in intersection crossing.
Studies in~\cite{perronnet2012cooperative,wu2012cooperative} focus on sequenced-based centralized CIM algorithms for safe trajectory planning where the RSU assigns right-of-way to CAVs.
The work in~\cite{li2020intersection} introduces a heuristic-based centralized CIM for CAVs that surpasses adaptive and static signalized intersections.
Some works~\cite{kowshik2011provable,gregoire2016hybrid} also suggest a hybrid CIM architecture to centrally allocate time slots using RSU and distributedly adjust CAV control for conflict-free crossing of CAVs through the intersection. 
However, given the earlier-mentioned constraints concerning centralized CIM systems due to RSU, distributed methods provide a cost-effective and scalable alternative, and thus are focused in this study as well.

\subsection{Distributed CIM for unsignalized intersections}
There is sparse literature on related distributed CIM 
for unsignalized intersections.
VanMiddlesworth~\etal~\cite{vanmiddlesworth2008replacing} use distributed CIM to navigate through 
unsignalized intersection,
however, it is only feasible for a low number of CAVs. Secondly, although the authors discuss possible CAV non-compliance consideration, however, they do not provide a mechanism to test it and confirm the robustness of the system.
Li~\etal~\cite{li2006cooperative} implement a distributed cooperative scheduling algorithm using 
trajectory planning for CAVs within a safe pattern to 
cross an intersection. 
Similarly, Katriniok~\etal~\cite{katriniok2019nonlinear} investigate a distributed trajectory planning based system where each CAV solves a non-linear Model Predictive Control (MPC) and shares its planned path with other CAVs.
Hassan~\etal~\cite{hassan2014fully} 
present a distributed heuristic-based algorithm for 
CIM
that outperforms the traditional First-in-First-Out (FIFO) approach at 
unsignalized intersections
in terms of average
intersection time
delay
in low and medium traffic, 
however, 
the performance highly deteriorates in high density traffic, hence the algorithm is not scalable,
as shown in our performance comparison later in the study.

\section{Driver Messenger System (DMS)}
DMS allows CAVs to utilize sensor data and V2X communication to share application-specific Over-the-Air (OTA) DIMs between Host Vehicle (HV) and Target Remote Vehicles (RVs).
For this study, HV is the term used to refer to one particular reference CAV out of all the CAVs in the scenario. Since DMS and all its encompassing applications such as the SC-I application in this study are distributed, every CAV in the scenario runs the system within its respective OBU. However, we just fix and denote a particular CAV as the HV and visualize the working of the application and the underlying algorithm, AROW in this case, from its perspective. 
DMS is essentially divided
into
four main components, which have been previously explained in detail by the authors \cite{shah2022enabling}. Relevant details of these components specific to the application in focus are mentioned in the following.

\subsection{Local Object Map}
This is the first module within DMS and it creates and updates a local object map to localize the HV and the nearby RVs around it. 
For this paper, the authors only utilize GPS and V2X communication to create and update the local object map and localize the CAVs.
It is assumed that all participating vehicles in any scenario are connected and that there are no communication losses due to congestion or bad resource allocation.

\subsection{Application Detection}
This module is tasked with utilizing the updated local object map and deciding if the HV is currently in one or more of the pre-defined applications within DMS. Once identified, DMS executes the corresponding application thread for the next application-specific tasks.
In this study, we assume a four-way allway SC-I where each of the lanes connecting to the SC-I is a singular lane. However, the logic presented throughout in this study along with the proposed AROW algorithm can be applied to any kind of SC-I. Figure \ref{fig:sci_det} shows an example of the SC-I used for this study. 

Since only GPS and V2X communication in the form of Basic Safety Messages (BSMs) are utilized in the creation of the local object map, this study uses a pre-configured map setting with an SC-I whose location is already known by all the participating CAVs as ground-truth. Therefore, as CAVs approach the SC-I, they can utilize the known ground truth location of the SC-I to determine their distance $(dist)$ from the SC-I and thus, the stop sign. The application detection block considers 
HV
to be within an SC-I application and trigger its application block as long as (\ref{eq_detthresh}) below is satisfied:
\begin{equation}
    dist \leq det_{thresh}
    \label{eq_detthresh}
\end{equation}
where
$dist$ 
is the distance of 
HV
from the stop sign in its lane and $det_{thresh}$ is the SC-I application detecting distance threshold. $det_{thresh}$ is one of the configurable parameters of SC-I as explained in later sections.


Another objective for 
HV
within the application detection block upon approaching an SC-I is to detect whether it is the leading CAV in its lane to reach the stop sign or if it is following another CAV in front of it.
As shown in Figure \ref{fig:sci_det}, 
HV
can be described in a typical SC-I scenario as shown in (\ref{eq_vehs}) below:
\begin{equation}
    HV = \begin{dcases*}
    v_L\in V_L & if leading, $dist \leq det_{thresh}$\\
    v_F\in V_F & if following, $dist \leq det_{thresh}$
    \end{dcases*}
\label{eq_vehs}
\end{equation}
where $v_L$ is the leading CAV in a particular lane while approaching an SC-I and it satisfies (\ref{eq_detthresh}) and $V_L$ is the current updated set of all leading CAVs in a SC-I. On the other hand, CAV $v_F$ is following another CAV in front while approaching an SC-I while also satisfying (\ref{eq_detthresh}) and $V_F$ is the set of all following CAVs in an SC-I. 
If HV satisfies (\ref{eq_detthresh}) and has a preceding and leading CAV, it is still classified in set $V_F$. Since AROW arbitration is only performed among CAVs in set $V_L$, it is sufficient to group all non-leading CAVs satisfying (\ref{eq_detthresh}) to set $V_F$.
Finally, CAV $v_O$ in set $V_O$ may or may not be the leading CAV however it does not satisfy (\ref{eq_detthresh}), and thus does not qualify for SC-I application.
As explained in the next section, finding the state of 
HV
and its proximity to an upcoming SC-I as per (\ref{eq_vehs}) is significant to determine its priority in the proposed AROW algorithm, especially in scenarios where there are queues of CAVs approaching an SC-I.

\subsection{Application Block}
Once the application detection block identifies that the HV is approaching an SC-I (\ref{eq_detthresh}), DMS 
within HV
triggers the application execution block of SC-I. This block contains the proposed AROW algorithm whose main goal is to identify fellow CAVs at the SC-I through the local object map and negotiate SC-I crossing turns of all CAVs in consideration by a series of DIMs. It should be noted that all CAVs in this study are assumed to be equipped with DMS and thus the same AROW algorithm is going to be executed for all CAVs approaching an SC-I.

\subsection{V2X Communication}
This block is responsible for transmitting and receiving DIMs to and from the HV, respectively. 
DIMs are application-specific and their structure and content varies for every specific application within DMS.
The details of DIM structure and encoding/decoding procedures are not the focus of this paper and thus are not included.



\section{Automated Right-of-Way (AROW) Algorithm }
\label{section:AROWsection}
Table \ref{table:arow_dims_acks} serves as a reference for all variables used within AROW and their descriptions throughout this study. 
AROW algorithm is executed inside the application module of SC-I within DMS where CAVs can utilize DIMs to assign priorities to cross an SC-I and reduce ambiguity. The SC-I application module within a HV is executed as soon as it satisfies (\ref{eq_detthresh}) and it can be referred to as belonging to either $V_L$ or $V_F$ as per (\ref{eq_vehs}).
On an abstract level, AROW considers the current total competing vehicles ($V_T$) at an SC-I such as 
shown in (\ref{eq_vt_vl_vf_union}),
\begin{equation}
    V_T = V_L \cup V_F
    \label{eq_vt_vl_vf_union}
\end{equation}
and it operates by assigning one of the $V_L$ as the arbitrator CAV ($v_{arb}$) such that $\mathit{v_{arb} \in V_L}$ can assign priorities in the form of turns to all CAVs within $V_L$. 
One of the main contributions of AROW is that it is equipped to assign distinct crossing turns such that there is no duplicity and ambiguity. Secondly, it is also capable of countering situations where some participating CAVs in an SC-I scenario can turn out to be non-compliant with the rules of AROW and turns assigned by $v_{arb}$. 


AROW comprises of several stages given the involved complexity and the task of maintaining resolution among multiple independent CAVs. 
Each stage within AROW involves transceiving of stage-specific DIMs and/or acknowledgment (ACK) signals. 

%
\begin{table}[t]
\centering
\caption{\small{AROW Variables and Descriptions.}}
\begin{tabular}{l  r}
\multicolumn{2}{c}{\small{Stages}}\\
\hline
\hline
$S0$   
&\text{Not in SC-I}\\
$S1$                &\text{SC-I Discovery}\\
$S2_1$                & \text{$v_{arb}$ Selection}\\
$S2_2$                &\text{$v_{arb}$ Announcement}\\
$S2_3$                & \text{$v_{arb}$ Acceptance}\\
$S3_1$                &\text{ Turn Scheduling)}\\
$S3_2$                & \text{Turn Acceptance}\\
$S4_1$ & \text{Exit SC-I + Assign New $v_{arb}$}\\
$S4_2$ & \text{Exit SC-I}\\
$SW$ & \text{Wait for current AROW completion}\\
\hline
\hline
\multicolumn{2}{c}{\small{DIMs and ACKs}}\\
\hline
\hline
AROW1   
&\text{SC-I introduction ($S1$)}\\
AROW2                &\text{$v_{arb}$ announcement ($S2_2$)}\\
ACK2                & \text{Acknowledgment of AROW2 ($S2_3$)}\\
AROW3                &\text{$v_{arb}$ turn scheduling ($S3_1$)}\\
ACK3                & \text{Acknowledgment of AROW3 ($S3_2$)}\\
AROW4\textsubscript{1}                &\text{ $v_{arb}$ in-turn SC-I exit + assign new $v_{arb}$ ($S4_1$)}\\
AROW4\textsubscript{2}                & \text{In-turn SC-I exit ($S4_2$)}\\
AROW5 & \text{Out-of-turn SC-I exit ($S2_2$,$S2_3$,$S3_1$,$S3_2$)}\\
AROW\textsubscript{wait} & \text{Request CAVs in $V_S$ to wait for current AROW}\\
\hline
\hline
\multicolumn{2}{c}{\small{Clock Constraints/Durations}}\\
\hline
\hline
$T1$   
&\text{Timeout for $S1$}\\
$T2$                &\text{Timeout for $S2_2$/$S2_3$}\\
$T3$                & \text{Timeout for $S3_1$/$S3_2$}\\
$T$\textsubscript{turn}                &\text{Timeout for $S4_1$/$S4_2$}\\
$T$\textsubscript{wait}                & \text{Timeout for $SW$}\\
\hline
\hline
\multicolumn{2}{c}{\small{Sets / Model Parameters}}\\
\hline
\hline
$V_L$   
&\text{Set of leading CAVs satisfying (\ref{eq_detthresh})}\\
$V_F$                &\text{Set of following CAVs satisfying (\ref{eq_detthresh})}\\
$V_O$                & \text{Set of CAVs not satisfying (\ref{eq_detthresh})}\\
$V_T$                &\text{Union set of $V_L$ and $V_F$}\\
$V_{P}$                & \text{Set of primary CAVs part of current AROW}\\
$V_{S}$                & \text{Set of secondary CAVs not part of current AROW}\\
HV                    & \text{Fixed reference CAV}\\
$dist$                & \text{Distance of HV to SC-I}\\
$det_{thresh}$                &\text{SC-I detection distance threshold}\\
$v_{arb}$                & \text{Arbitrator CAV for current AROW}\\
$NC_{stages}$ & \text{Non-compliance stages ($S2_2$, $S2_3$, $S3_1$, $S3_2$)}\\
$NC_{prob}$ & \text{Non-compliance probability}\\
$AROW_{count}$ & \text{Counter for consecutive AROW repetitions}\\
$AROW_{thresh}$ & \text{Threshold for $AROW_{count}$}\\
\hline
\hline
\multicolumn{2}{c}{\small{Heuristics}}\\
\hline
\hline
$H_1$                &\text{Heuristic in $S2_1$ for $v_{arb}$ selection of current AROW}\\
$H_2$                &\text{Heuristic in $S3_1$ for SC-I turn assignment}\\
$H_3$                & \text{Heuristic in $S4_1$ for $v_{arb}$ selection of next AROW}\\
\hline
\multicolumn{2}{c}{\small{Timed Automaton}}\\
\hline
\hline
$\mathcal{M}$                &\text{Automaton}\\
$S$                &\text{Set of locations}\\
$S^0$                & \text{Set of initial start location(s) ($S^0 = S0$)}\\
$A$                & \text{Set of input symbols}\\
$T$                & \text{Singular clock used in $\mathcal{M}$}\\
$X$                & \text{Set of clocks ($X=T$)}\\
$\mathit{\Phi(T)}$                & \text{Set of clock constraints on $T$}\\
$I(s)$                & \text{Mapping between $s\in S$ and $\mathit{\Phi(T)}$}\\
$E$                & \text{Set of switches between locations in $S$}\\
$\lambda$                & \text{Set of clock(s) to be reset during switch ($\lambda \subseteq T$)}\\
$N$                & \text{Transition system of $\mathcal{M}$}\\
\hline
\end{tabular}
\label{table:arow_dims_acks}
\end{table}
%

\subsection{Stage $0$ $(S0)$ - Not in SC-I}
This is the idle stage of AROW and the HV stays in this stage as long as it does not satisfy (\ref{eq_detthresh}). 

\subsection{Stage $1$ $(S1$) - SC-I Discovery}
This is the first step in the execution of AROW where the HV satisfies (\ref{eq_detthresh}) and broadcasts a discovery DIM known as AROW1.
On a high-level, in addition to HV's basic safety and identification data, the contents of this DIM can be considered to highlight the detection of SC-I by the HV and the initiation of AROW algorithm within its DMS. Also, this DIM includes information pertaining to whether currently 
$\text{HV} \in V_L \vert \vert \text{ HV} \in V_F$.
Just like the HV, any CAV starting its AROW process can be assumed to broadcast a similar AROW1 DIM.
After broadcasting AROW1, the HV waits for a pre-configured
duration/clock-constraint 
$T1$ to receive any 
DIMs from other CAVs in $V_T$. The two types of receiving DIMs under focus in this stage are AROW1s from CAVs that have also just initiated AROW algorithm or AROW\textsubscript{wait} from $v_{arb}$.

Towards the end of 
duration
$T1$, if the HV does not receive any DIM, then it can be considered that there are no fellow CAVs at the SC-I besides the HV itself. This triggers DMS to effectively terminate AROW and the HV is instructed by its DMS to cross the SC-I without AROW and switch to $S0$. 
On the other hand, if HV receives AROW1(s) from other CAVs but no AROW\textsubscript{wait}, then this signifies that there is no $v_{arb}$ present at SC-I but there are fellow competing CAVs that arrived at the SC-I at similar times as that of the HV, and that the AROW process should proceed into next stage, i.e. $S2_1$. Finally, if the HV receives AROW\textsubscript{wait} from $v_{arb}$, this signals that there is a current AROW process going on where $v_{arb}$ is arbitrating and negotiating SC-I crossing turns among a group of CAVs ($V_P$) that arrived at the SC-I at least $T1$ time before the HV. 
AROW\textsubscript{wait} is sent by $v_{arb}$ to any CAV 
broadcasting AROW1 to signal it to wait for the current AROW process to end before 
this CAV
can participate for its own turn assignment. As detailed in the next sub-section, as soon as $v_{arb}$ is assigned in $S2_1$ and until all its tasks in AROW are accomplished, $v_{arb}$ maintains two updating sets in which it divides the CAVs from $V_T$ such that:
\begin{itemize}
    \item $V_P$: Set containing primary CAVs that are part of the current AROW procedure. These are always the leading CAVs. 
    \item $V_S$: Set containing secondary CAVs that are instructed to wait for the current AROW procedure to finish before they can be considered for arbitration. These can either be following CAVs or leading CAVs that arrive during an ongoing AROW.
\end{itemize}

At any time within AROW, arbitration is always performed among CAVs that are present in $V_P$.
Therefore, if HV receives AROW\textsubscript{wait} from $v_{arb}$, then this indicates that HV needs to move to 
AROW Stage Wait ($SW$)
and wait for the current arbitration to end. 
In this case, HV is also added to $V_S$ as it is not one of the CAVs considered for the current round of arbitration.
Further details of $SW$ and its operation is described later in Section \ref{Section SW}.


 

\subsection{Stage $2_1$ $(S2_1)$ - $v_{arb}$ Selection}
In the case that HV receives at least one AROW1 from a fellow CAV but no AROW\textsubscript{wait} from $v_{arb}$, 
AROW enters $S2_1$ in which a single CAV from $V_L$ is chosen as $v_{arb}$ for the remainder of the AROW process until $v_{arb}$ crosses the SC-I on its assigned turn.
Based on a fixed heuristic $H_1$ within DMS of every CAV in set $V_L$, one of the CAVs is chosen as $v_{arb}$.
To ensure that all participating CAVs in $V_L$ choose the same $v_{arb}$ and there is no duplication, there is a single heuristic stored within DMS of all CAVs. For this study, the heuristic $H_1$ is as follows:
\begin{itemize}
    \item $H_1$: The last arriving CAV denoted from the timestamp at which it entered SC-I application by satisfying (\ref{eq_detthresh}) and initiated its AROW, is chosen as $v_{arb}$. 
    If two or more CAVs arrive at the same timestamp, 
    then their unique CAV identification numbers
    such as Medium Access Control (MAC) addresses 
    can be utilized as tie breaker and the CAV with an alphanumeric character containing the lowest ASCII value at the first index with differing characters by process of elimination can be chosen as $v_{arb}$.
    Depending on user preference, alternative approaches for tie-breaking can also be employed such as unique CAV Vehicle Identification Number (VIN), etc.
    
\end{itemize}
Following this same $H_1$, it can be guaranteed that all CAVs in $V_L$ will choose the same CAV as $v_{arb}$.
It should be noted that no DIM transmission or reception occurs in $S2_1$ and that it is simply a computational stage where a $v_{arb}$ is chosen for the rest of the current arbitration round. 

Particular attention needs to be given to the fact that during this process of $v_{arb}$ selection, only CAVs that are considered are the ones that were a part of $V_L$ up until the end of duration $T1$. In other words, any new CAVs that entered the SC-I application 
after the current AROW process considering CAVs in $V_L$ has already moved on from $S1$ would not be considered in the current arbitration selection process even if it is now a part of $V_L$, since priority is given to CAVs that arrived at similar times in $S1$ and have already passed to the next stage. 

Whoever is chosen as the $v_{arb}$ in this stage as per $H_1$, the first task that it needs to perform is to divide up CAVs in $V_T$ into $V_P$ and $V_S$, as mentioned in the previous sub-section. All the CAVs as part of the current arbitrator process in $S2_1$ are included in $V_P$ whereas all the CAVs in $V_F$ are added to $V_S$. 
Also, if any CAV enters SC-I and its $S1$ as a $v_L$ or $v_F$ while the current AROW process is going on in any stage after $S1$, it would be directly entered in set $V_S$ by $v_{arb}$.
At every timestamp throughout AROW, whenever any CAV enters or leaves SC-I, $v_{arb}$ always updates both $V_P$ and $V_S$. 

Finally, during $S2_1$, if the HV is chosen to be the $v_{arb}$ as per $H_1$, then it goes into $S2_2$ and all the other CAVs in $V_P$ proceed to their $S2_3$,
and vice-versa.


\begin{figure}[t]
\centerline{\includegraphics[trim=5 0 5 0,clip,width=.48\textwidth]{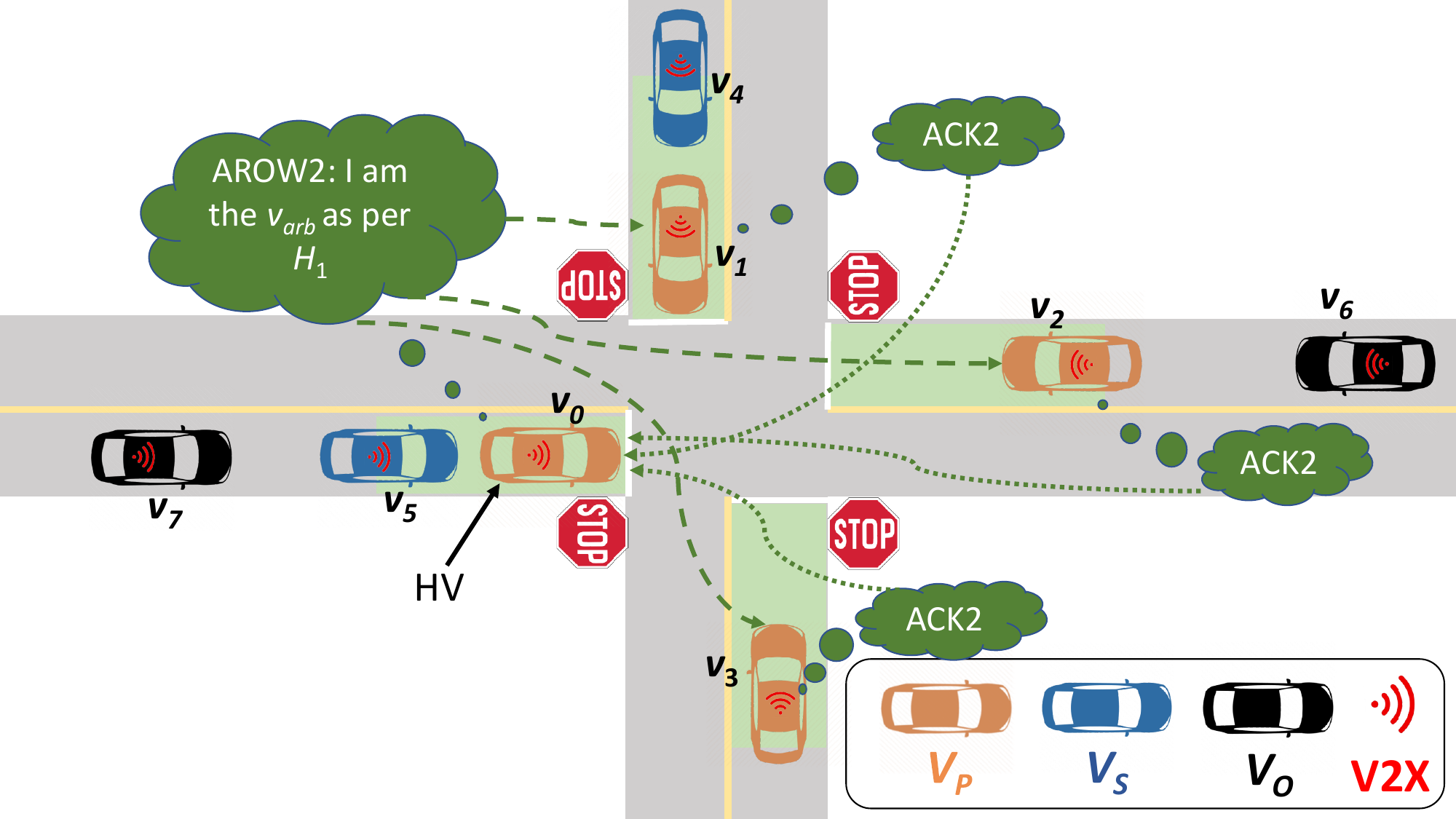}}
\caption{\small{Example scenario during HV's $v_{arb}$ announcement in stage $S2_2$. HV had previously broadcasted its AROW2 DIM and the other competing CAVs in $V_P$ are now sending their ACK2 signals.}}
\label{fig:arow_s22}
\end{figure}

\subsection{Stage $2_2$ $(S2_2)$ - $v_{arb}$ Announcement}
In $S2_2$, the HV broadcasts AROW2 DIM in which it informs all the CAVs in $V_P$ that it has been chosen as $v_{arb}$. After broadcasting AROW2, the HV waits for 
duration
$T2$ to receive acknowledgment signal ACK2 from CAVs in $V_P$ as an acceptance of HV as $v_{arb}$.
Figure \ref{fig:arow_s22} shows an example of AROW in $S2_2$ with HV as $v_{arb}$.
After the completion of $T2$, if HV receives ACK2 from all $V_P$, then HV moves to $S3_1$, whereas the non-arbitrator CAVs in $V_P$ move to $S3_2$.
Alternatively, if there is non-compliance where one or more CAVs other than HV in $V_P$ leave the SC-I prematurely, the other compliant CAVs in $V_P$ along with any new CAVs in $V_L$ that are not yet part of $V_P$ can re-start AROW by going back to $S1$. The number of times the compliant CAVs can consecutively repeat a particular round of AROW in face of non-compliance is determined by the pre-configured $AROW_{thresh}$, as shown in the equation below: 
\begin{equation}
    AROW_{count} \leq AROW_{thresh}
    \label{eq_arowthresh}
\end{equation}
where $AROW_{count}$ is the counter to keep track of the consecutive repetitions.
As soon as (\ref{eq_arowthresh}) is not satisfied, DMS within all compliant CAVs in $V_P$ switches to $S0$ and the CAVs can cross the SC-I manually without the assistance of AROW. 



\subsection{Stage $2_3$ $(S2_3)$ - $v_{arb}$ Acceptance}
In $S2_3$, since the HV is not chosen as $v_{arb}$, it waits for $T2$ to receive AROW2 DIM from $v_{arb}$, and upon reception, it needs to send an ACK2. Towards the end of $T2$, if all CAVs in $V_P$ along with HV are compliant, then AROW moves HV into $S3_2$ along with other CAVs in $V_P$ that are not $v_{arb}$. At the same time, AROW within $v_{arb}$ moves it from $S2_2$ into $S3_1$ as explained in the previous subsection. 
However, if there is any non-compliance by a CAV in $V_P$ or $v_{arb}$ itself, then the AROW within HV and other compliant CAVs in $V_P$ requires to break the current AROW process and move to either $S1$ or $S0$ depending on (\ref{eq_arowthresh}), as explained in the previous subsection.

\begin{figure}[t]
\centerline{\includegraphics[trim=5 0 5 0,clip,width=.48\textwidth]{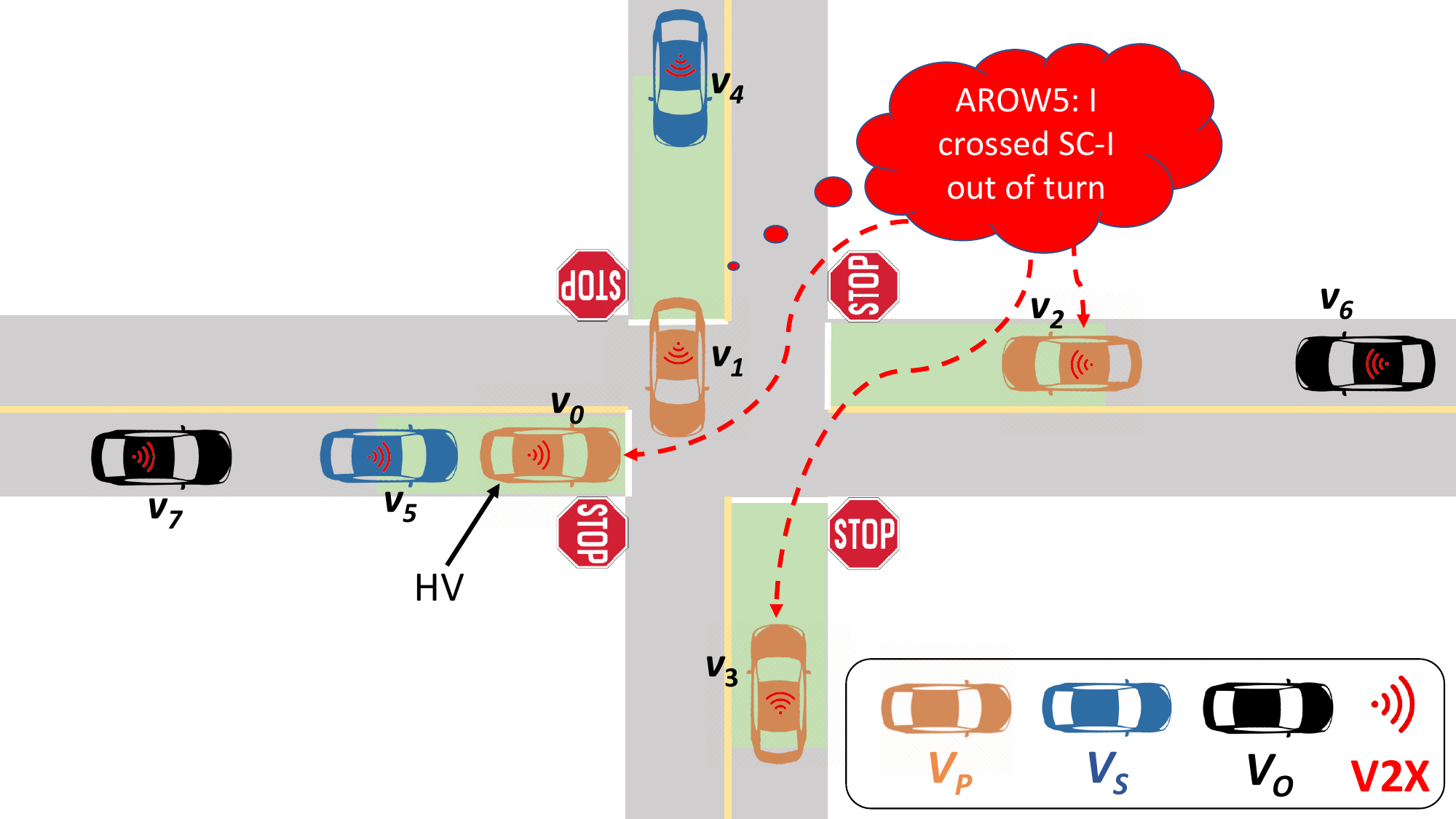}}
\caption{\small{Non-compliance scenario in AROW with HV as $v_{arb}$. One of the CAVs in $V_P$ has become non-compliant and prematurely exited the SC-I upon which the DMS within it has broadcasted the AROW5 DIM.}}
\label{fig:arow_nc}
\end{figure}

\begin{figure}[t]
\centerline{\includegraphics[trim=5 0 5 0,clip,width=.48\textwidth]{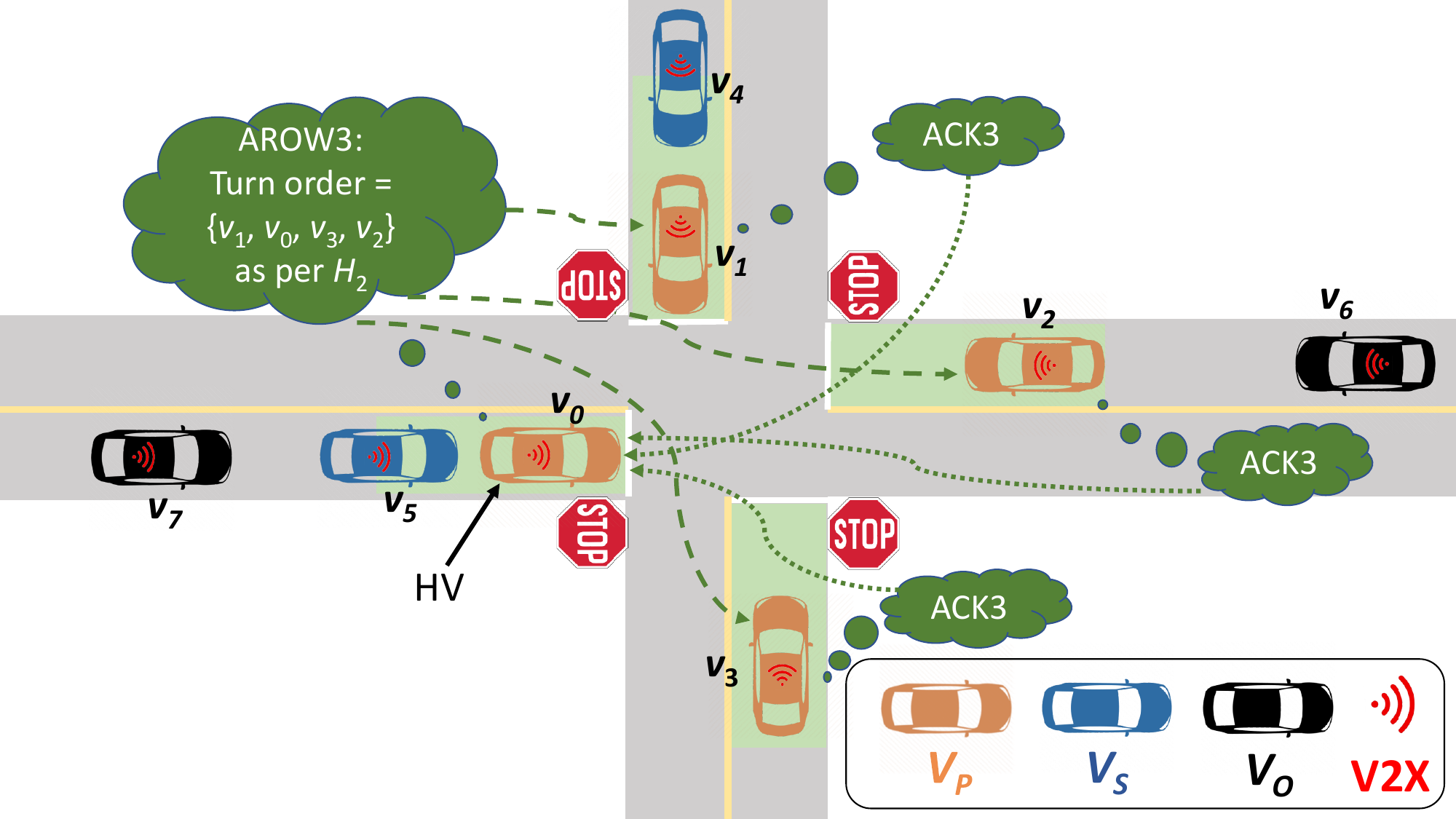}}
\caption{\small{Example scenario during HV's $v_{arb}$ turn assignment in stage $S3_1$. HV had previously broadcasted its AROW3 DIM and the other competing CAVs in $V_P$ are now sending their ACK3 signals.}}
\label{fig:arow_s31}
\end{figure}

\subsection{Non-Compliance}
\label{subsection:NC}
Non-compliance is a condition within AROW that can occur when a CAV is in any of four stages defined as $NC_{stages} = \{S2_2, S2_3, S3_1, S3_2\}$. A CAV in $V_P$ is said to be non-compliant whenever it crosses an SC-I prior to its assigned turn. Whenever it does, the out-of-turn CAV broadcasts an AROW5 exit DIM. This DIM is used by $v_{arb}$ and other CAVs in $V_P$ to identify a non-compliant CAV. 
Figure \ref{fig:arow_nc} shows an example of non-compliance where one of the CAVs in $V_P$ decides to leave during $S2_2$ where turns have not even been assigned yet. As soon as AROW within that CAV detects the untimely exit, it broadcasts AROW5.
Usually, DMS would utilize a camera to identify if it has already passed the SC-I to broadcast AROW5, however, since we only utilize V2X as the data source for this study, the ground truth data of the SC-I is utilized for the purpose of identifying an exit from an SC-I and broadcasting AROW5 DIM.
In this study, non-compliance can occur to any CAV in $V_P$ during AROW including $v_{arb}$. The only exception to this is the HV itself. In other words, we always assume that the HV is always fully compliant.
To evaluate the impact of non-compliance, it is probabilistically applied ($NC_{prob}$) to each CAV independently when it is in each of the four stages in $NC_{stages}$. 
\subsection{Stage $3_1$ $(S3_1)$ - Turn Scheduling}
HV can arrive in $S3_1$ from either the parent stage $S2_2$ or $SW$. The details of switching from the latter parent stage are explained later under the subsection of $SW$. 
In the case of parent stage $S2_2$, if the HV receives ACK2 signal from all CAVs in $V_P$ after its selection as $v_{arb}$, HV moves into $S3_1$ while the other CAVs in $V_P$ move to $S3_2$. This is the stage where the HV as $v_{arb}$ assigns turns to all CAVs in $V_P$ to cross the SC-I based on a fixed pre-configured heuristic $H_2$. 
Figure \ref{fig:arow_s31} shows an example of AROW in this stage. 
The main goal behind choosing any heuristic is to ensure that distinct turns are assigned to each CAV and that there is no duplication. The heuristic $H_2$ used for this study is defined as follows:
\begin{itemize}
    \item $H_2$: $v_{arb}$ assigns turns based on the arrival timestamp of every CAV in $V_P$ and prioritizes CAVs that arrived earlier at the SC-I. In the situation where more than one CAV carries identical arrival timestamp, $v_{arb}$ can randomly assign turns among them.
\end{itemize}
As it can be observed in $H_2$ description, the tie-breaking is performed by random selection. Based on user preference, it can be changed to some other heuristic too. For instance, the tie-breaking rules can also be governed by using the unique CAV identification numbers as done for $H_1$. It should be noted that this notion furthers the proposed argument that AROW can incorporate any heuristic $H_2$ for turn-assignment as per user preferences, as long as all CAVs contain this same heuristic. 

The turn assignment information is enclosed within AROW3 DIM and broadcasted. Once received, the CAVs in $V_P$ are required to reply with an ACK3 acknowledgment to confirm the corresponding turn assignment. 
After broadcasting AROW3, HV waits for 
duration
$T3$.
Towards the end of $T3$, if the HV receives ACK3 signals from all relevant CAVs, then it moves towards $S4_1$. On the other hand, if there is non-compliance by any CAV, then depending on (\ref{eq_arowthresh}), HV either moves to $S1$ to restart AROW or to $S0$ where it disables AROW to manually pass the SC-I.

\subsection{Stage $3_2$ $(S3_2)$ - Turn Acceptance}
After passing $S2_3$, HV moves into $S3_2$ where, similar to that in $S2_3$, it waits for $v_{arb}$ to send a AROW3 DIM that contains the turn assignment for the HV among other CAVs in $V_P$. Upon receiving AROW3, HV sends an ACK3 signal to acknowledge the reception. Once $T3$ is elapsed and there is no non-compliance, the HV moves into $S4_2$. 
However, just like in $S2_3$, if any CAV in $V_P$ or $v_{arb}$ becomes non-compliant, then HV moves into $S1$ or $S0$ depending on (\ref{eq_arowthresh}).

\begin{figure}[t]
\centerline{\includegraphics[trim=5 0 5 0,clip,width=.48\textwidth]{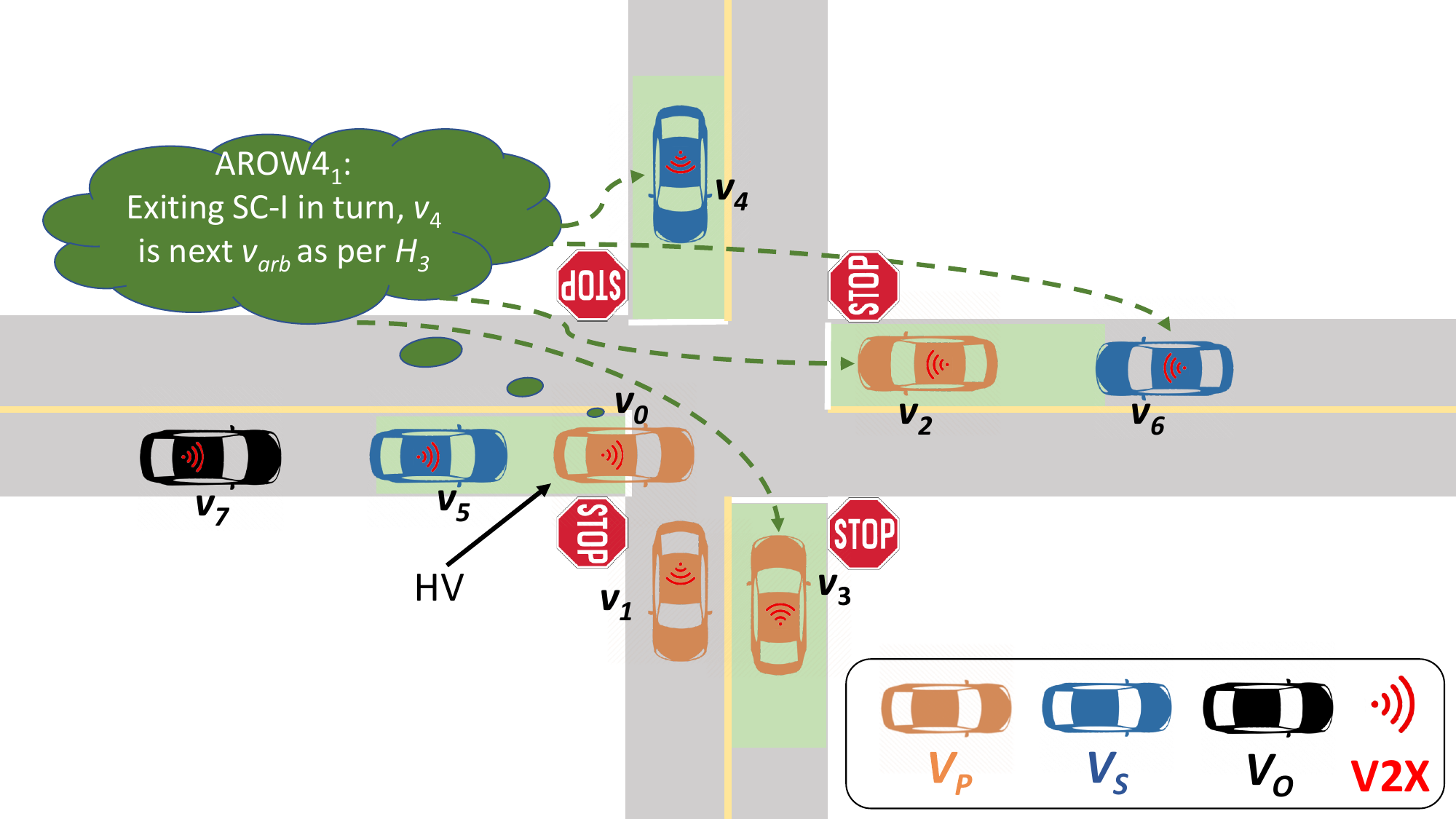}}
\caption{\small{Example scenario during HV's $v_{arb}$ exiting SC-I on its turn and broadcasting AROW4\textsubscript{1} DIM in stage $S4_1$ and assigning new $v_{arb}$ for next AROW round.}}
\label{fig:arow_s41}
\end{figure}

\subsection{Stage $4_1$ $(S4_1)$ - Exit SC-I + Assign New $v_{arb}$}
$S4_1$ is the ultimate stage for HV as $v_{arb}$ in AROW, as shown in Figure \ref{fig:arow_s41}. By the time the HV arrives in this stage and the other CAVs in $V_P$ arrive in the corresponding $S4_2$, the turns have already been assigned to all involved CAVs along with the HV. HV stays in this stage until the arrival of its turn to cross the SC-I. Upon exiting the SC-I, the HV needs to perform one last task  in its role as $v_{arb}$ i.e. to choose a new $v_{arb}$ as per a pre-defined heuristic $H_3$. For simplicity, this study uses $H_1$ as the heuristic for choosing new $v_{arb}$, i.e. ($\mathit{H_3 = H_1}$). If there are any CAVs in $V_S$ waiting at the SC-I for the current AROW procedure involving HV and $V_P$ to terminate, then HV as $v_{arb}$ needs to select one of CAVs in $V_S$ to become the new $v_{arb}$.
After selecting new $v_{arb}$, HV crosses the SC-I, and broadcasts AROW4\textsubscript{1} DIM indicating to all other CAVs in $V_T$ that it is exiting SC-I along with the information pertaining to new $v_{arb}$.
Finally, AROW within HV moves to $S0$ as the idle stage.

\subsection{Stage $4_2$ $(S4_2)$ - Exit SC-I}
Similar to $S4_1$, HV waits for its turn to pass the SC-I. Once it does, it broadcasts a final AROW4\textsubscript{2} DIM, indicating to CAVs in $V_T$ that it is leaving the SC-I, and then its AROW moves to $S0$.



\subsection{Stage Wait $(SW)$ - Wait for current AROW completion}
\label{Section SW}
If HV arrives at SC-I by satisfying (\ref{eq_detthresh}) with an ongoing AROW arbitration round, then the $v_{arb}$ instructs HV via AROW\textsubscript{wait} to wait in $SW$ as one of $V_S$. If the current AROW concludes successfully without any non-compliance, then depending on the set $V_S$, the HV can either move to $S0$, $S3_1$, or $S3_2$. If HV is the only CAV in $V_S$, then AROW is not required and HV can manually pass the SC-I, thus HV moves to $S0$. In the case of presence of other CAVs in $V_S$, HV can move to $S3_1$ if $v_{arb}$ of the concluding AROW has chosen HV to be the new $v_{arb}$ via AROW4\textsubscript{1} DIM, otherwise, it transitions to $S3_2$. 
The process of assigning new $v_{arb}$ assists in optimizing AROW since it avoids the need for CAVs to utilize AROW from the beginning stage of $S1$.
Finally, if there is non-compliance among competing CAVs in $V_P$ while HV is in $SW$ and (\ref{eq_arowthresh}) is not satisfied which means that all CAVs in $V_P$ collectively eliminate AROW and manually pass SC-I, then HV along with others CAVs in $V_S$ move to $S1$ to initiate AROW from the beginning. 

\begin{figure*}
\centering
\includegraphics[trim=520 250 220 0, clip,width=250mm]
{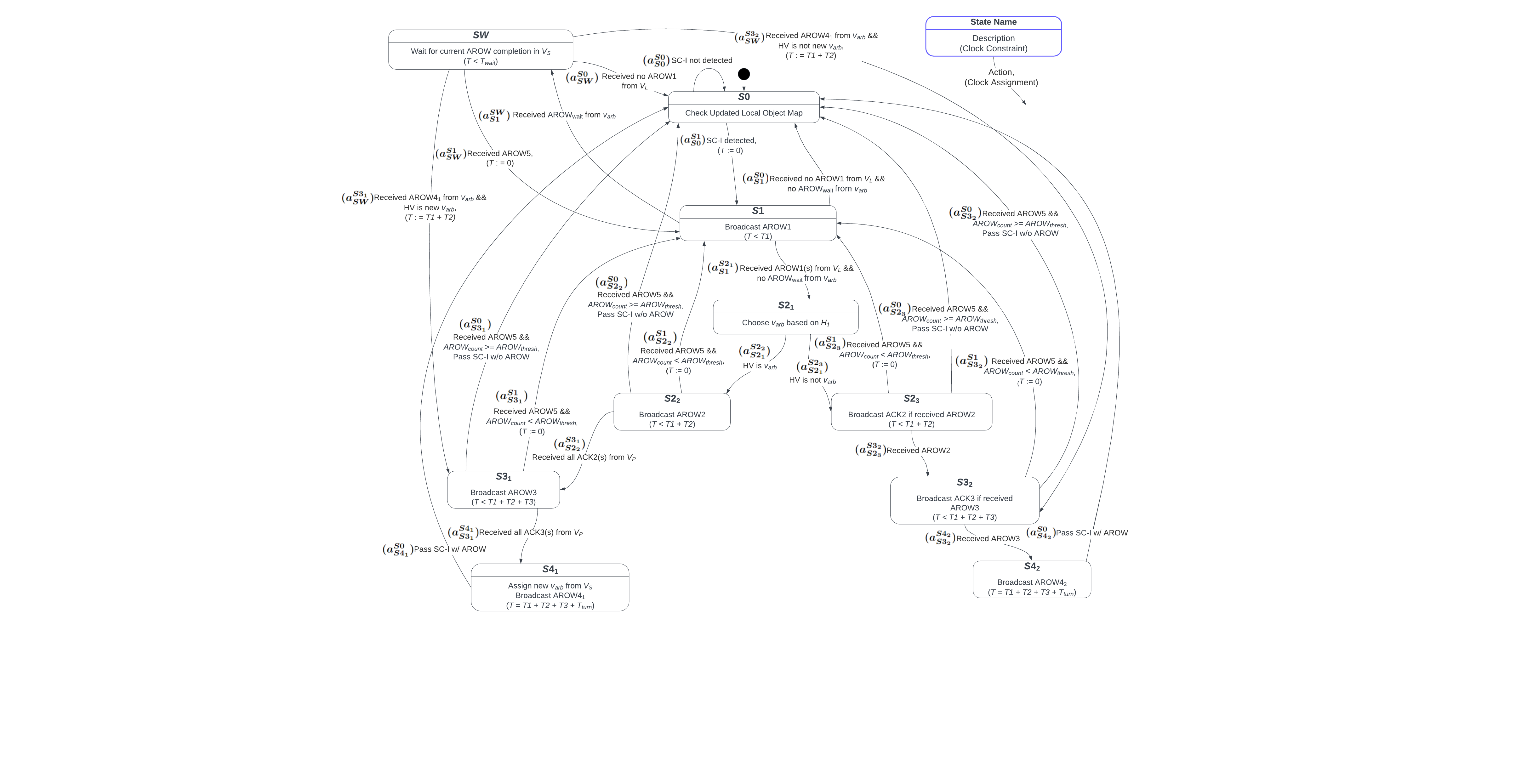}
\caption{\small{AROW algorithm expressed as a timed automaton.}}
\label{fig:arow_sm}
\end{figure*}

%
%

\section{AROW Implementation as a Timed Automaton}
\label{Section: Timed automaton}
This section provides details into the earlier-introduced stages within AROW and shows their functionality as distinguishable states. 
AROW can be modeled as a timed automaton \cite{alur1994theory}, \cite{alur1999timed} that allows it to be within a particular state at any specific time due to clock constraints.
Figure \ref{fig:arow_sm} represents AROW as a timed automaton. The automaton consists of nodes called locations and edges known as switches.
Formally, the automaton can be described as a tuple $\mathcal{M} = \langle S, S^0, A, X, I, E\rangle$ where
\begin{itemize}
    \item $S$ is a finite set of all locations within $\mathcal{M}$ and it can be defined as
    \begin{equation}
        S = \{SW, S0, S1, S2_1, S2_2, S2_3, S3_1, S3_2, S4_1, S4_2\} 
    \end{equation}
    Each location in $S$ defines a particular AROW stage which has been explained in detail in the previous section. Figure \ref{fig:arow_sm} provides a description highlight for each location.
    
    \item $S^0 \subseteq S$ is a set of initial locations. For AROW system, 
    \begin{equation}
        S^0 = \{S0\}    
    \end{equation}
    This is also the location where AROW is typically deactivated which means that whenever HV is in this location, it is not approaching a nearby SC-I.
    
    \item $A$ is a finite set of input symbols,
    \begin{equation}
        A = \{ a_{s}^{s'} \mid s,s' \in S\}
        \label{eq-inputsymbol}
    \end{equation}
    All the possible $s$ and $s'$ for the input symbols can be viewed in Figure \ref{fig:arow_sm}.

    \item $X$ is a finite set of clocks. The transition system for AROW only utilizes a singular clock $T$, hence $X = T$. 

    \item $I$ is a mapping that assigns locations within $\mathcal{M}$ with clock constraints in set 
    $\mathit{\Phi(T)}$ as shown in (\ref{eq:phi_T}). These are the same constraints that were previously defined as durations in Section \ref{section:AROWsection}.
    \begin{equation}
    \label{eq:phi_T}
    \begin{aligned}
        \mathit{\Phi(T)} = \{\,
        T1, 
        T1+T2, 
        T1+T2+T3,\\ 
        T1+T2+T3+T\textsubscript{turn},
        T\textsubscript{wait}\}    
    \end{aligned}    
    \end{equation}
    These constraints are included in specific locations within the description blocks as observable in Figure \ref{fig:arow_sm}.
   $I$ maps these constraints from 
   $\mathit{\Phi(T)}$ 
   on select locations within $\mathcal{M}$ as mentioned below:

    \begin{itemize}
        \item $S1: T < T1$
        \item $S2\textsubscript{2}, S2\textsubscript{3}: T < T1+T2$
        \item $S3\textsubscript{1}, S3\textsubscript{2}: T < T1+T2+T3$
        \item $S4\textsubscript{1}, S4\textsubscript{2}: T < T1+T2+T3+T\textsubscript{turn}$
        \item $SW: T < T\textsubscript{wait}$
    \end{itemize}
    All locations within $\mathcal{M}$ not mentioned in the mapping above can be considered as not possessing any clock constraint.
    
    \item
    $\mathit{E \subseteq S \textsc{ x } A \textsc{ x } \Phi(T) \textsc{ x } S }$
    is a set of switches between locations within $\mathcal{M}$. A switch 
    $\langle s, a_{s}^{s'}, \lambda, s' \rangle$
    represents a transition edge from location $s$ to $s'$ over a symbol 
    $a_{s}^{s'}$. 
    A switch can contain the set 
    $\lambda \subseteq T$ 
    that allows 
    $T$ 
    to be reset to a particular value with this switch.
    There exist some switches within $\mathcal{M}$ in which the clock $T$ is reset, 
    hence, non-empty $\lambda$.
    All such instances where
    $e_{s}^{s'} \subseteq E$ switches contain a clock $T$ reset are mentioned in (\ref{eq:T_reset}):
    \begin{equation}
        \small
        \label{eq:T_reset}
        T = 
        \begin{cases}
         0 & \text{if } s = S1, s' \in \{S0, S2_2, S2_3, S3_1, S3_2, SW\},\\

         T1 + T2 & \text{if } s = SW, s' \in \{S3_1, S3_2\} 
        \end{cases}
    \end{equation}
    Every $s$ and $s'$ in $e_{s}^{s'}$ mentioned above corresponds to the switch containing an input symbol $a_{s}^{s'} \in A$ as in (\ref{eq-inputsymbol}).
    \end{itemize}
\begin{table}[t]
\centering
\caption{\small{SUMO simulation parameters.}}
\begin{tabular}{l  r}
\hline
\hline
Experiment Duration   
&50000\text{s}\\
No. of CAVs ($n_{veh}$)               &4, 8, 16\\
Car-Following Model                & Krauss~\cite{krauss1997metastable}\\
Maximum Vehicle Speed                   & 70 \text{m/s}\\
Vehicle Acceleration
    & 2.6\text{m/s\textsuperscript{2}}\\
Vehicle Deceleration
    & 4.5\text{m/s\textsuperscript{2}}\\
Sigma (Driver Imperfection)     &0.5\\
minGap (Empty space after leader)  &2.5\text{m}\\
CAV length & 5\text{m}\\
No. of lanes/road segment
    & 1\\
\hline
\end{tabular}
\label{table:configs_sumo}
\end{table}

    $N$ is the transition system of $\mathcal{M}$. State of $N$ can be described as a pair
    $(s,T)$ 
    where $s \in S$ denotes the current location and 
    $T$
    is the clock such that it
    satisfies the clock constraint mapped to $s$ by $I(s)$. 
    A transition 
    can be described in two ways:
    \begin{itemize}
        \item In the case of self-transition where a transition state returns to itself due to clock constraint,
        \begin{equation}
            N_{s}^{s'} : (s, T) \xrightarrow[]{\delta} (s, T+\delta),
        \end{equation}
        where the clock increment $\delta$ is such that 
        $T+\delta$ 
        satisfies the clock constraint set by
        $I(s)$
        mapping. Also, the source and destination location is 
        $s$ 
        signifying the self-transition.
        \item In the case of transition switch from one location 
        $s$ 
        to another 
        $s'$ 
        based on $e_{s}^{s'} \in E$,
        \begin{equation}
            N_{s}^{s'} : (s, T) \xrightarrow[]{a_{s}^{s'}} (s', T[\lambda := y]),
        \end{equation}
        where 
        $a_{s}^{s'}$
        is part of the transition edge 
        $e_{s}^{s'}$.
        Also,
        $s$ and $s'$
        are the source and destination locations, respectively. $y$ represents the reset of the clock upon reaching the destination location. 
        
    \end{itemize}
This transition system will be repeatedly utilized for analysis in the next sections. The focus will be on transitions caused by location switches and will generally be mentioned as
$N_{s}^{s'}$ for simplicity.

\section{Experimental Results}

\subsection{Simulation and AROW Configuration Parameters}
To test AROW and provide experimental details regarding its functionality,
the transition system $\mathcal{M}$ defined in the last section is modeled and tested using the Simulation of Urban Mobility (SUMO) traffic simulator \cite{SUMO2018}.
The simulation parameters specific to SUMO are mentioned in Table \ref{table:configs_sumo}. These parameters can be configured as per user preference. 
The road topology used for this study considers a four-way allway SC-I attached to four single-lane road segments. Each end of the four road segments from SC-I is about 51.2m in length and connected to a circular round-about to facilitate CAVs to drive in an endless loop and repeatedly pass through the SC-I in the center for the entirety of the simulation.
In order for the AROW algorithm to interact in real-time with SUMO simulation, the authors utilize Traffic Control Interface (TraCI) \cite{wegener2008traci}. 
To test a variety of use-cases, the study utilizes different number of CAVs, i.e. $n_{veh}$ = 4, 8, or 16. Given the road topology, these values of $n_{veh}$ can be considered as low, medium, and high density traffic, respectively.

 

Table \ref{table:configs_arow} shows the values assigned to the configurable parameters of AROW for this study. 
The $det_{thresh}$ is chosen to be 10m with aim to: (i) allow the CAVs to initiate the process of AROW in advance so that generally the CAV is assigned its turn to leave SC-I by the time it arrives at the stop sign, provided there is no non-compliance, and (ii) to allow up to two CAVs per lane to be included in the AROW process given the CAV length mentioned in Table \ref{table:configs_sumo}.
%
\begin{table}[t]
\centering
\caption{\small{AROW configurable parameters.}}
\begin{tabular}{l  r}
\hline
\hline
$det_{thresh}$   
&10\text{m}\\
BSM Periodicity                &\{100,600\}\text{ms}~\cite{saej3161}\\
Resource Reservation Interval &100\text{ms}\cite{toghi2018multiple}\\
$T1$
    & 2\text{s}\\
$T2$                & 2\text{s}\\
$T3$                   & 2\text{s}\\
$NC_{prob}$                   & 0, 0.1, 0.25\\
$AROW_{thresh}$
    & 2\\
\hline
\end{tabular}
\label{table:configs_arow}
\end{table}
On the other hand, $AROW_{thresh} = 2$ to provide compliant CAVs with at most two consecutive chances to use AROW to eliminate ambiguity in the face of non-compliance. $AROW_{thresh}$ should not be set to a really high value since it can then unfairly penalize compliant CAVs for following AROW rules since they will end up spending more time waiting to pass the SC-I.

As explained earlier, $T1$ is the clock constraint used by a CAV during discovery stage $S1$ of AROW. 
$T1$ for this study is set to 2s to allow realistic chance for one or more CAVs to initiate AROW and these CAVs can then enter in an arbitration as mentioned in the AROW logic explained in the previous sections. 
$T2$ is the clock constraint used to control how long a CAV stays in $S2\textsubscript{2}$ or $S2\textsubscript{3}$ while undergoing AROW process. 
The experimental value of $T2$ is chosen to be 2s for CAVs in $S2_2$ and $S2_3$. The realistic value for $T2$ can be further reduced and optimized based on the inherent communication latencies of the V2X technology in use. This optimization is the subject of a future study related to this topic. 
The same logic is applied to choosing the value for $T3$. 

\begin{table}[b]
\setlength{\tabcolsep}{1.8pt} 
\caption{\small{Raw state frequencies from simulations in Figure \ref{fig:arow_simcompletion} with full compliance ($NC_{prob} = 0$).}}
\begin{center}
\begin{tabular}{|c|c|c|c|c|c|c|c|c|c|c|c|}
\hline
\boldmath$n_{veh}$ & \boldmath$SW$ & \boldmath$S0$ & \boldmath$S1$ & \boldmath$S2_1$ & \boldmath$S2_2$ & \boldmath$S2_3$ & \boldmath$S3_1$ & \boldmath$S3_2$ & \boldmath$S4_1$ & \boldmath$S4_2$ & \textbf{Total}\\
\hline
\multirow{2}{*}{\textbf{4}} & \textbf{814} & \textbf{1577} & \textbf{1577} & \textbf{256} & \textbf{130} & \textbf{126} & \textbf{211} & \textbf{201} & \textbf{211} & \textbf{201} & \textbf{5304}\\\cline{2-12} & 
15.35 & 29.73 & 29.73 & 4.83 & 2.45 & 2.38 & 3.98 & 3.79 & 3.98 & 3.79
& 100\%\\
\hline
\multirow{2}{*}{\textbf{8}} & \textbf{1146} & \textbf{1159} & \textbf{1159} & \textbf{12} & \textbf{4} & \textbf{8} & \textbf{400} & \textbf{675} & \textbf{400} & \textbf{675} & \textbf{5638}\\\cline{2-12} &
20.33 & 20.56 & 20.56 & 0.21 & 0.07 & 0.14 & 7.09 & 11.97 & 7.09 & 11.97 & 100\%\\
\hline
\multirow{2}{*}{\textbf{16}} & \textbf{587} & \textbf{588} & \textbf{588} & \textbf{1} & \textbf{0} & \textbf{1} & \textbf{167} & \textbf{421} & \textbf{167} & \textbf{421} & \textbf{2941}\\\cline{2-12} &
19.96 & 19.99 & 19.99 & 0.03 & 0 & 0.03 & 5.68 & 14.31 & 5.68 & 14.31 & 100\%\\
\hline
\end{tabular}
\end{center}
\label{table:State Count raw}
\end{table}


\subsection{Analysis and Results}
\subsubsection{AROW Performance under Full Compliance ($NC_{prob}=0$)}
\begin{figure}[t]
\centering
\includegraphics[trim=0 0 0 0,width=0.47\textwidth]{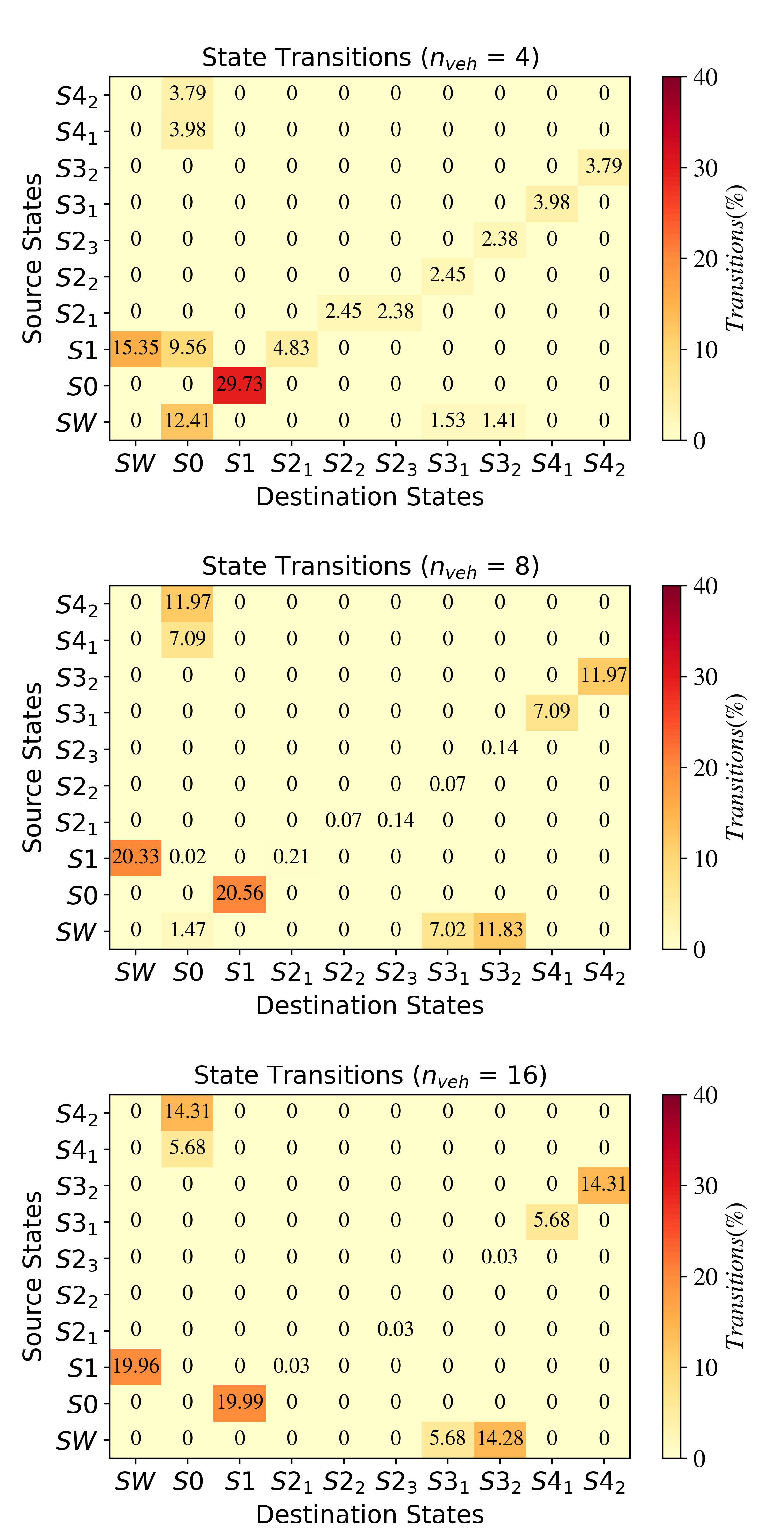}
\caption{\small{State transition comparison for simulations with full compliance ($NC_{prob}$ = 0).}}
\label{fig:arow_simcompletion}
\end{figure}

The first goal using simulations is to prove the completeness of the logic of AROW. To achieve this, the simulations using AROW with parameters mentioned in Tables \ref{table:configs_sumo} and \ref{table:configs_arow} are run and each state's frequency and transitions are independently logged throughout the simulation duration.
For this set of simulations, all the participating CAVs are set to be fully compliant, i.e. $\mathit{NC_{prob}}$ = 0.
It should be noted that the terms `state frequency' and `state transitions' are used instead of `location frequency' and `location transition' to specify that these values have been collected over time in a simulation whenever the clock constraints for each transition are fulfilled as mentioned earlier in the description of the timed automaton $\mathcal{M}$ of AROW. 
The results of these simulations are shown in Figure \ref{fig:arow_simcompletion} and Table \ref{table:State Count raw}. 
Figure \ref{fig:arow_simcompletion} shows the state transitions matrices represented as heat maps and based on the transition system $N$ of AROW. 
The x-axis and y-axis denote the destination and source locations, respectively. 
In addition to Figure \ref{fig:arow_simcompletion}, Table \ref{table:State Count raw} provides the state frequency in terms of raw numbers and percentages for each location.

For each of the three simulations, Table \ref{table:State Count raw} shows state frequency of each location and the transition matrix provides the outgoing and incoming transition switches from that location, respectively. 
For example, for $\mathit{n_{veh} = }$ 4, $S3_1$ has a percentage frequency of 3.98\% and the transition matrix clarifies this frequency by showing that $S3_1$ has two incoming transition switches 
$N_{SW}^{S3_1}$
and 
$N_{S2_2}^{S3_1}$, 
with frequencies of 1.53\% and 2.45\%, respectively, that add up to 3.98\%, equal to state frequency of $S3_1$ as shown in Table \ref{table:State Count raw}. Additionally, the transition matrix also shows one outgoing transition switch 
$N_{S3_1}^{S4_1}$
from $S3_1$ 
worth 3.98\% which is again equal to the incoming transitions to $S3_1$ and also its percentage frequency. 
The equality between incoming and outgoing transition switches can be observed for all locations in all traffic scenarios, except for minuscule differences due to rounding in percentages. All the transitions in the matrices can also be observed to be in line with those in the timed automaton $\mathcal{M}$ and use switches $E$ corresponding to input symbols $A$ in Figure \ref{fig:arow_sm}, thereby ensuring that the experimental operation is corresponding to the AROW theory. Finally, all transitions within each matrix along with the state frequency percentages in Table \ref{table:State Count raw} add up to 100\%, barring rounding errors.

Comparing the three simulations, it can be first observed that as the value of $n_{veh}$ increases from 4 to 8, the total raw frequency increases by a small margin as shown in Table \ref{table:State Count raw}. This is mainly due to the fact that this higher number of CAVs in the scenario increases the frequency of HV's transition to $SW$ upon its arrival at the SC-I in focus. Furthermore, as $n_{veh}$ increases to 16, there is a sharp reduction in the total raw frequency which is caused by a reduction in the count of $S0$ and $S1$. 
This is mainly because as $n_{veh}$ increases, there is a higher congestion at the SC-I in focus. Thus, the HV spends more time to get through SC-I and hence, fewer transitions switches of 
$N_{S0}^{S1}$
from $S0$ to $S1$ in the given simulation duration.
The percentage equality of $S0$ and $S1$ in every simulation also attests to the 100\% 
$N_{S0}^{S1}$
transition as shown in the graph $\mathcal{M}$ in Figure \ref{fig:arow_sm}.

From $S1$, the transition 
$N_{S1}^{S0}$
to $S0$ occurs when the HV is unaccompanied by any fellow competing CAV at the SC-I, and in this case, the HV passes SC-I without AROW. As expected, the matrix for $n_{veh} = 4$ shows 
$N_{S0}^{S1}$
transition frequency of 9.56\%, which goes to almost 0\% in the case of $n_{veh} = 8, 16$ since with more CAVs, there is a lesser probability of HV approaching the SC-I in focus without any competition. On the other hand, the transition 
$N_{S1}^{SW}$
occurs when HV initiates AROW in the middle of an ongoing AROW round among CAVs in $V_P$. In this case, $v_{arb}$ instructs HV to wait in $SW$. It can be seen in both the percentage state frequency and matrices that the frequency of $N_{S1}^{SW}$ is higher for $n_{veh} = 8, 16$ because as $n_{veh}$ increases, there is a higher probability that HV would arrive at SC-I in focus and initiate in the presence of other CAVs.

One more transition possibility from $S1$ is
$N_{S1}^{S2_1}$.
$N_{S1}^{S2_1}$
occurs when HV is accompanied by other competing CAVs in $V_L$ in SC-I and there is no $v_{arb}$ present. In this case, the CAVs along with HV need to follow AROW and elect a $v_{arb}$. Again, as $n_{veh}$ increases, there is a higher probability of a $v_{arb}$ already present upon the arrival of HV, thus the frequency of 
$N_{S1}^{S2_1}$.
reduces. This is why 
$N_{S1}^{S2_1}$
frequency for $n_{veh}=4$ is 4.83\% whereas it drops to just 0.21\% and 0.03\% for $n_{veh}=8$ and $n_{veh}=16$, respectively. 
From $S2_1$, the only transitions are
$N_{S2_1}^{S2_2}$
and
$N_{S2_1}^{S2_3}$
to $S2_2$ and $S2_3$, respectively. These transitions can be seen in the transition matrices and for the same afore-mentioned reason of a higher frequency of 
$N_{S1}^{S2_1}$
for $n_{veh} = 4$, we can also observe a higher frequency of 
$N_{S2_1}^{S2_2}$
and
$N_{S2_1}^{S2_3}$
for $n_{veh} = 4$ that sum to the total state frequency of
$N_{S1}^{S2_1}$.
Similarly, the transitions from $S2_2$ and $S2_3$ are 
$N_{S2_2}^{S3_1}$
and
$N_{S2_3}^{S3_2}$
to $S3_1$ and $S3_2$, respectively, and carry a higher frequency in the case of $n_{veh}=4$ as compared to other scenarios. It should be noted that since $NC_{prob} = 0$ in this set of three simulations, the transitions
$N_{S2_2}^{S0}$, $N_{S2_2}^{S1}$, $N_{S2_3}^{S0}$, and $N_{S2_3}^{S1}$
are all 0\%.

For $SW$, the transitions are 
$N_{SW}^{S0}$, $N_{SW}^{S1}$, $N_{SW}^{S3_1}$, and $N_{SW}^{S3_2}$. 
$N_{SW}^{S0}$ 
occurs in the cases when HV is ready to switch from $SW$ upon the conclusion of current AROW round of other CAVs in $V_P$, however, there is no other waiting CAV besides HV, thus AROW is not required for HV. As expected, this transition majorly occurs in $n_{veh}=4$ scenario with less CAVs (12.41\%), and the other scenarios with higher CAVs, the frequency of
$N_{SW}^{S0}$ 
reduces rapidly. The transition 
$N_{SW}^{S1}$ 
is 0\% for all scenarios here since all CAVs are fully compliant ($NC_{prob} = 0$). The transitions 
$N_{SW}^{S3_1}$ 
and
$N_{SW}^{S3_2}$ 
occur in the cases when HV, along with other waiting CAVs in $V_S$ is ready to switch out from $SW$ upon the conclusion of the current AROW round and the $v_{arb}$ has chosen one of the CAVs from $V_P$ to become the new $v_{arb}$. If HV is chosen as the new $v_{arb}$, then it follows 
$N_{SW}^{S3_1}$, 
otherwise it uses
$N_{SW}^{S3_2}$ 
as per AROW rules. As shown in the Figure \ref{fig:arow_simcompletion} transition matrices, these transitions contain higher frequencies as $n_{veh}$ increases, simply due to the earlier-stated argument that a higher number of competing CAVs creates a higher probability of CAVs in $SW$.

Finally, from $S3_1$ and $S3_2$, the only valid transitions in the case of $NC_{prob} = 0$ are
$N_{S3_1}^{S4_1}$ 
and
$N_{S3_2}^{S4_2}$, 
respectively. For the same reason, the transitions
$N_{S3_1}^{S1}$, $N_{S3_1}^{S0}$, $N_{S3_2}^{S1}$, and $N_{S3_2}^{S0}$ 
are 0\%. With the increase of $n_{veh}$, there are more instances of competing CAVs with HV, and thus the percentage of 
$N_{S3_1}^{S4_1}$ 
and 
$N_{S3_2}^{S4_2}$ 
also rises. 
Once CAVs successfully conclude AROW in $S4_1$ or $S4_2$, AROW is transitioned back to $S0$ using transitions
$N_{S4_1}^{S0}$ 
and
$N_{S4_2}^{S0}$, 
respectively.

\subsubsection{AROW Performance under Non-Compliance Scenarios}
\paragraph{Theoretical analysis of Non-Compliance}
To increase robustness, AROW considers the occurrence of non-compliance and allows compliant CAVs to still reduce ambiguity and safely pass SC-I. To show the impact of non-compliance, this study randomly assigns non-compliance probability, i.e., $NC_{prob}$, to each CAV except HV  every time they engage in AROW. There are two values used for experimentation purposes, i.e. $NC_{prob}$ = 0.1, or 0.25.
Even higher values can be chosen, but for the purposes of this paper, these values are sufficient to show the collective impact of $NC_{prob}$ on AROW.
In theory, $NC_{prob}$ is designed to be independently applied to each CAV during each of the stages in $NC_{stages}$. 

To show the collective impact of $NC_{prob}$ on a typical round of AROW, we use a metric $P(\exists NC)$ which is the probability that at least one CAV becomes non-compliant throughout a singular round of AROW.
Using the law of total probability $P(X) = \sum_{n}P(X \cap Y_{n})$,
we calculate $P(\exists NC)$ below:
\begin{equation}
\label{eq:atleast1NC_arow}
    P(\exists NC) =  \sum_{j}^{n_{CV}}\sum_{m}^{j}\sum_{i}^{NC_{stages}} P(HV_{i})P(NC_{m}|HV_{i})P(CV_{j}),
\end{equation}
where $P(HV_i)$ is the probability that HV is in stage $i$, $P(NC_{m}|HV_{i})$ is the conditional probability that there are $m$ non-compliant CAVs when HV is in stage $i$, and $P(CV_j)$ is the probability of HV in the current AROW round with $j$ competing CAVs. $P(CV_j)$ uses the outermost sum where $CV \in V_P$ and $n_{CV}$ is the total number of competing CAVs to the HV during a singular round of AROW. Considering the experimental topology for this study where competing CAVs can be on any of the single-lane roads of the SC-I, $n_{CV}$ can be ranged as $1 \leq j \leq 3$.
For every $j$ in the outermost sum, the inner sum aggregates $P(NC_{m}|HV_{i})$ in range  $1 \leq m \leq j$. Finally, over every $m$, the inner-most sum aggregates $P(HV_i)$ where $i$ is every location defined in $NC_{stages}$, i.e. $NC_{stages} = \{S2_2, S2_3, S3_1, S3_2\}$.
As can be observed in (\ref{eq:atleast1NC_arow}), the probability $P(\exists NC)$ is calculated from the perspective of the HV due to the easiness of calculations. For example, if HV is in $S2_2$, we can already infer that other CAVs in $V_P$ will be in $S2_3$, or if HV is in $S2_3$, we can then infer that one of the CAVs in $V_P$ other than HV has been chosen as $v_{arb}$ and has proceeded to $S2_2$, whereas remaining CAVs have moved to $S2_3$ along with HV. The same logic can be applied for $HV$ being in $S3_1$ or $S3_2$. 

$P(CV_j)$ is non-deterministic since it is not possible to predict the specific number of competing CAVs to HV at a specific time, and thus in this experiment, we rely on simulations to estimate it.
On the other hand, when calculating probability $P(NC_{m}|HV_{i})$ and given the set $NC_{stages}$, it can be observed that $P(NC_{m}|HV_{S2_{2}}) = P(NC_{m}|HV_{S2_{3}})$ and similarly, $P(NC_{m}|HV_{S3_{1}}) = P(NC_{m}|HV_{S3_{2}})$. On a high-level, $P(NC_{m}|HV_{i})$ can be defined as 
a binomial distribution as shown below in (\ref{eq:P(NCm|HVi)}):

\begin{equation}
\label{eq:P(NCm|HVi)}
\small
    P(NC_{m}|HV_{i}) =
        \binom{j}{m} * P(NC | HV_i)^m*(1-P(NC | HV_i))^{(j - m)}
\end{equation}
where $\binom{j}{m} = \frac{j!}{m!(j-m)!}$ is the binomial coefficient representing the different combinations of $m$ non-compliant CAVs, $P(NC | HV_i) = NC_{prob}$, $P(NC | HV_i)^m$ is the probability of $m$ non-compliant CAVs, and $(1-P(NC | HV_i))^{(j - m)}$ is the probability of $(j - m)$ CAVs being compliant.
Finally, $P(HV_i)$ is calculated as:

\begin{equation}
\small
\label{eq:P(HV_i)}
    P(HV_{i}) = 
    \begin{cases}
    \begin{aligned}
        (P({N_{S1}^{S2_1}}^*) + P(N_{S1}^{SW})P(N_{SW}^{S1})P({N_{S1}^{S2_1}}^+))P(N_{S2_1}^{S2_2}) & \\\text{if } i = S2_2,\\
        (P({N_{S1}^{S2_1}}^*) + P(N_{S1}^{SW})P(N_{SW}^{S1})P({N_{S1}^{S2_1}}^+))P(N_{S2_1}^{S2_3}) & \\\text{if } i = S2_3,\\
        P(HV_{S2_2})P_C(S2_{2}) + P(N_{S1}^{SW})P(N_{SW}^{S3_1}) & \\\text{if } i = S3_1,\\
        P(HV_{S2_3})P_C(S2_{3}) + P(N_{S1}^{SW})P(N_{SW}^{S3_2}) & \\\text{if } i = S3_2\\
    \end{aligned}
    \end{cases}
\end{equation}

where 


\begin{equation}
    \label{eq:Pc}
    P_C(k) = (1-P(NC | HV_{k}))^{j}
\end{equation}

\begin{figure*}
\centering
\includegraphics[trim=100 0 100 0,width=0.75\textwidth]{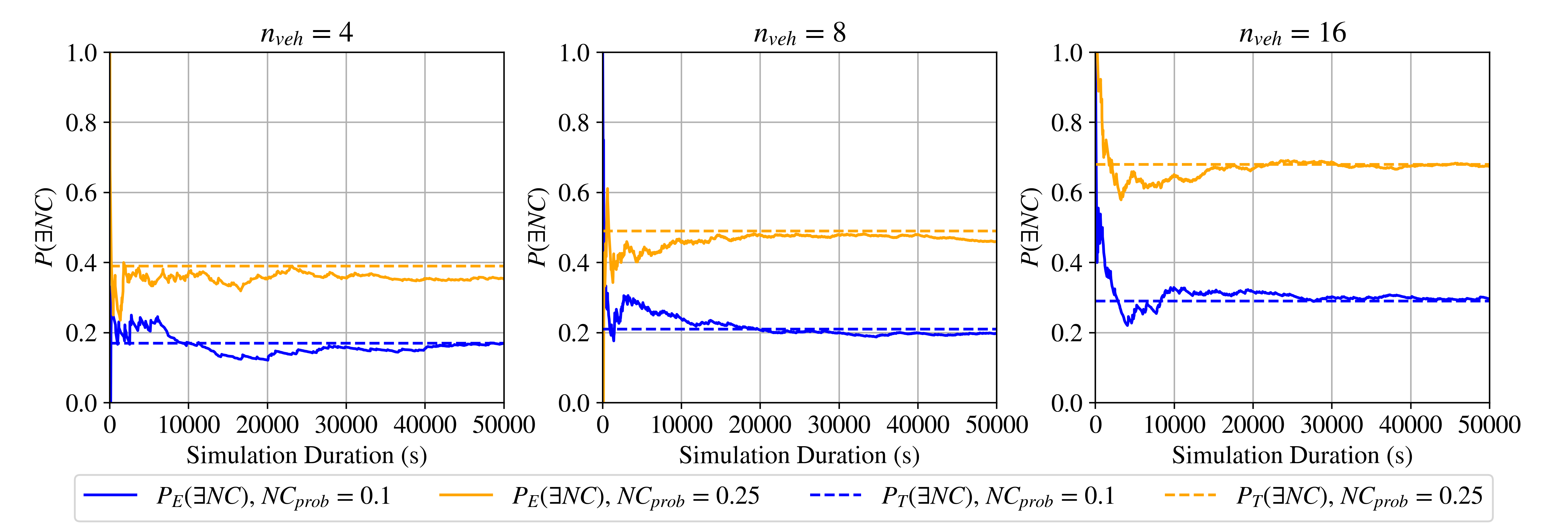}
\caption{\small{Comparison between $P_T(\exists NC)$ and $P_E(\exists NC)$.}}
\label{fig:atleast1NC_theo_exp}
\end{figure*}

As observable in (\ref{eq:P(HV_i)}), $P(HV_i)$ yields different values depending on the specific location and its parent locations. For instance, $P(HV_{S2_2})$ or the probability that HV arrives in $S2_2$ is dependent on either the direct transition
${N_{S1}^{S2_1}}^*$
or sequence of transitions 
$N_{S1}^{SW}$, $N_{SW}^{S1}$, and ${N_{S1}^{S2_1}}^+$.
After considering the possibilities of HV arriving at $S2_1$, the result is multiplied by 
$P(N_{S2_1}^{S2_2})$
to calculate $P(HV_{S2_2})$ 
It should be noted that $\|N_{S1}^{S2_1}\| = \|{N_{S1}^{S2_1}}^*\| + \|{N_{S1}^{S2_1}}^+\|$, and these transitions are separated out due to probability calculations. $P(HV_{S2_3})$ is also similarly calculated for HV in $S2_3$. Calculating $P(HV_{S3_1})$ and $P(HV_{S3_2})$ is comparatively more complex because in addition to arriving at $S3_1$ or $S3_2$ from $S2_2$ or $S2_3$ using transitions 
$N_{S2_2}^{S3_1}$
or
$N_{S2_3}^{S3_2}$,
respectively, HV can also arrive at $S3_1$ or $S3_2$ from the wait stage $SW$ using transitions 
$N_{SW}^{S3_1}$,
or
$N_{SW}^{S3_2}$,
respectively. 
In addition, in the cases of HV arriving at $S3_1$ or $S3_2$ from $S2_2$ or $S2_3$, it is also necessary for all competing CAVs to HV to be compliant in the parent locations $S2_2$ or $S2_3$, respectively. This can be calculated probabilistically using (\ref{eq:Pc}).

Given $j$ which is the number of competing CAVs to HV in the current iteration of (\ref{eq:atleast1NC_arow}), some probabilities in (\ref{eq:P(HV_i)}) such as 
$P(N_{S2_1}^{S2_2})$, $P(N_{S2_1}^{S2_3})$, $P(N_{SW}^{S3_1})$, and $P(N_{SW}^{S3_2})$
can be determined probabilistically, whereas other remaining probabilities such as
$P({N_{S1}^{S2_1}}^*)$, $P({N_{S1}^{S2_1}}^+)$, $P(N_{S1}^{SW})$, and $P(N_{SW}^{S1})$
are non-deterministic since they totally depend on the particular scenario. Finally, since we are calculating $P(\exists NC)$ where it is assumed that $1 \leq n_{CV} \leq 3$, thus 
$P({N_{S1}^{S2_1}}^+)$
is always assumed to be 1.

\begin{table}[t]
  \centering
  \caption{\small{Tabular comparison between $P_T(\exists NC)$ and $P_E(\exists NC)$.}}
  \begin{tabular}{|c|c|c|c|}
    \hline
     \multicolumn{1}{|c|}{\boldmath{$n_{veh}$}}  & \multicolumn{1}{c|}{\boldmath{$NC_{prob}$}} & \multicolumn{1}{c|}{\boldmath{$P_T(\exists NC)$}} & \multicolumn{1}{c|}{\boldmath{$P_E(\exists NC)$}}\\
     \hline
    \multirow{2}{*}{\textbf{4}} & \textbf{0.1} & 0.17 & 0.17 \\\cline{2-4}
    & \textbf{0.25} & 0.39 & 0.35\\
   \hline  
    \multirow{2}{*}{\textbf{8}} & \textbf{0.1} & 0.21 & 0.2 \\\cline{2-4}
    & \textbf{0.25} & 0.49 & 0.46\\
   \hline
       \multirow{2}{*}{\textbf{16}} & \textbf{0.1} & 0.29 & 0.29 \\\cline{2-4}
    & \textbf{0.25} & 0.68 & 0.68\\
   \hline
  \end{tabular}
  \label{table:P_T(NC) vs. P_E(NC)}
\end{table}

\begin{figure*}
\centering
\includegraphics[trim=100 0 100 0,width=0.75\textwidth]{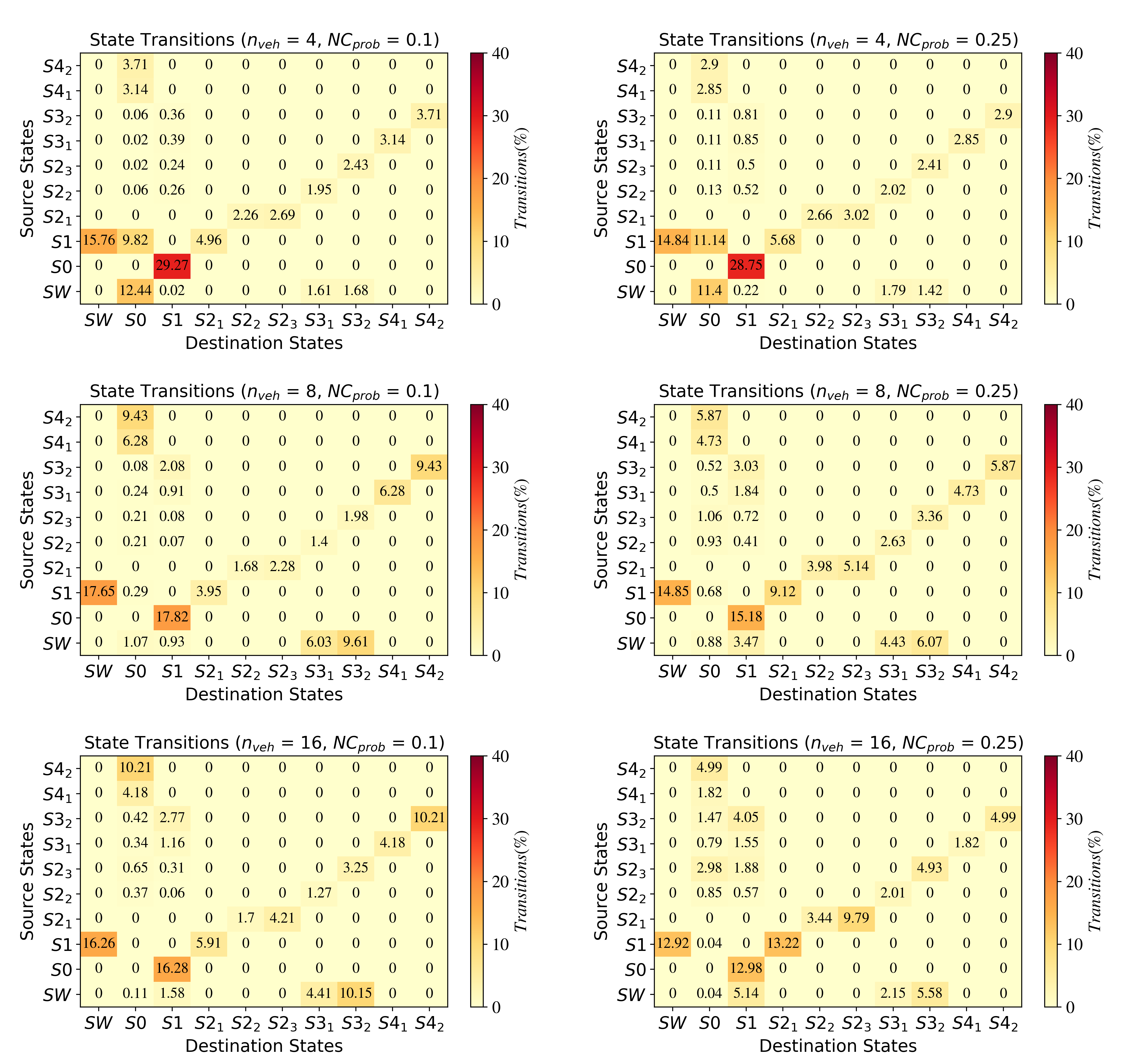}
\caption{\small{State transition comparison for non-compliance scenarios ($NC_{prob}$ = 0.1, $NC_{prob}$ = 0.25).}}
\label{fig:arow_nc_statetransitions}
\end{figure*}
\paragraph{Experimental Analysis of Non-Compliance}
Given the parameters in Tables \ref{table:configs_sumo} and \ref{table:configs_arow} and (\ref{eq:atleast1NC_arow}), (\ref{eq:P(NCm|HVi)}), (\ref{eq:P(HV_i)}), and (\ref{eq:Pc}), we run a total of six simulations using different values of $n_{veh}$ and $NC_{prob}$ and compare the theoretical and experimental $P(\exists NC)$, i.e. $P_E(\exists NC)$ and $P_T(\exists NC)$, as shown in Figure \ref{fig:atleast1NC_theo_exp}. It should be noted that the earlier-mentioned non-deterministic probabilities that are needed to calculate $P_T(\exists NC)$ are taken from the simulations for each scenario as known probabilities. As observable in all simulation results in Figure \ref{fig:atleast1NC_theo_exp}, $P_E(\exists NC)$ converges to $P_T(\exists NC)$ as the simulation moves ahead in time, which proves that AROW correctly considers the impact of $NC_{prob}$ and the experimental results match with that of the theoretical concept.
This also shows that the simulation duration and logging data size are sufficient to validate that the AROW algorithm has no deadlocks among a variation of scenarios.
Table \ref{table:P_T(NC) vs. P_E(NC)} points to the same convergence in tabular form 
as $P_T(\exists NC)$ and $P_E(\exists NC)$ are very similar to each other towards the end of the simulation duration. For all simulation scenarios, it can be generally viewed that as expected, an increase in $NC_{prob}$ results in an increase in the overall $P_T(\exists NC)$ and $P_E(\exists NC)$. Secondly, an increase in $n_{veh}$ also increases $P_T(\exists NC)$ and $P_E(\exists NC)$ for both values of $NC_{prob}$ since more $n_{veh}$ results in more competing CAVs with HV during every AROW round. As evident from Table \ref{table:Percentage of CVs}, as we increase $n_{veh}$ for both values of $NC_{prob}$, the percentage of AROW rounds involving HV with a higher number of competing CAVs rises. For instance, with simulations scenarios where $n_{veh}=4$, the highest percentage of AROW rounds for HV are with 1 competing CAV. For $n_{veh} = {8, 16}$, the highest percentage changes to 2 and 3 competing CAVs, respectively. Therefore, this explains why $P_T(\exists NC)$ and $P_E(\exists NC)$ increase with the increase in $n_{veh}$. 

\begin{table}[t]
  \centering
  \caption{\small{Percentages of competing CAVs ($CVs$) during AROW rounds for non-compliance scenarios ($NC_{prob}$ = 0.1, $NC_{prob}$ = 0.25).}}
  \begin{tabular}{|c|c|c|c|c|}
    \hline
    \multicolumn{2}{|c|}{} &\multicolumn{3}{c|}{\textbf{Percentage of \boldmath{$CVs$}}} \\
    \hline
     \multicolumn{1}{|c|}{\boldmath{$n_{veh}$}}  & \multicolumn{1}{c|}{\boldmath{$NC_{prob}$}} & \multicolumn{1}{c|}{\textbf{1}} & \multicolumn{1}{c|}{\textbf{2}}
     & \multicolumn{1}{c|}{\textbf{3}}\\
     \hline
    \multirow{2}{*}{\textbf{4}} & \textbf{0.1} & 88 & 10.7  & 1.3\\\cline{2-5}
    & \textbf{0.25} & 90.6 & 8.8 & 0.6\\
   \hline  
    \multirow{2}{*}{\textbf{8}} & \textbf{0.1} & 25.1 & 57.7  & 17.2\\\cline{2-5}
    & \textbf{0.25} & 34.4 & 54.8 & 10.8\\
   \hline
       \multirow{2}{*}{\textbf{16}} & \textbf{0.1} & 0.9 & 29.3  & 69.8\\\cline{2-5}
    & \textbf{0.25} & 3.3 & 29.7 & 67\\
   \hline
  \end{tabular}
  \label{table:Percentage of CVs}
\end{table}

After establishing the fact that the theoretical impact of $NC_{prob}$ matches with that of the experimental results, we analyze the effect of $NC_{prob}$ on AROW performance in terms of the state transitions. 
For this reason, Figure \ref{fig:arow_nc_statetransitions} displays the state transition matrices in the form of heat maps for different values of $n_{veh}$ and $NC_{prob}$.
Most transitions and trends in the heat maps in Figures \ref{fig:arow_simcompletion} and \ref{fig:arow_nc_statetransitions} are similar except for the transitions involving $NC_{stages}$.
The first major difference is that the heat maps in Figure \ref{fig:arow_nc_statetransitions} contain non-zero transitions in
$N_{S2_2}^{S0}$, $N_{S2_2}^{S1}$,
$N_{S2_3}^{S0}$, $N_{S2_3}^{S1}$,
$N_{S3_1}^{S0}$, $N_{S3_1}^{S1}$,
$N_{S3_2}^{S0}$, and $N_{S3_2}^{S1}$,
due to $NC_{prob}$, as opposed to the same transitions which were all 0\% in the heat maps in Figure \ref{fig:arow_simcompletion} due to full compliance. 

It has been mentioned that in the case of non-zero $NC_{prob}$, AROW within every compliant competing CAV utilizes $AROW_{thresh}$ so that in the cases of non-compliance by some other CAV(s), it does not need to withdraw from using AROW altogether and transition to $S0$ but instead re-attempt AROW until (\ref{eq_arowthresh}) is being satisfied. This is the reason why the transitions 
$N_{S2_1}^{S2_2}$
and
$N_{S2_1}^{S2_3}$
yield higher percentages with $NC_{prob}$ = 0.25 than $NC_{prob}$ = 0.1, which in turn yields higher percentages than $NC_{prob}$ = 0.
The same reasoning is also responsible for the increase in percentages of 
$N_{S2_2}^{S3_1}$
and
$N_{S2_3}^{S3_2}$
in the cases of $NC_{prob}$ = 0.25 than $NC_{prob}$ = 0.1.

Next, for any $n_{veh}$ in Figure \ref{fig:arow_nc_statetransitions},  as $NC_{prob}$ increases, so does the percentage of non-compliance related transitions from any of the $NC_{stages}$. For example, it can be observed that 
$N_{S2_2}^{S1}$
is higher for $NC_{prob}= $0.25 as compare to that for $NC_{prob}= $0.1. The same is the case with 
$N_{S2_2}^{S0}$, $N_{S2_3}^{S1}$, $N_{S2_3}^{S0}$, $N_{S3_1}^{S1}$, $N_{S3_1}^{S0}$, $N_{S3_2}^{S1}$, and $N_{S3_2}^{S0}$.
This can also be understood by analyzing that for any $n_{veh}$, the total non-compliance percentage by taking the sum of
$N_{S2_2}^{S1}$, $N_{S2_2}^{S0}$ $N_{S2_3}^{S1}$ $N_{S2_3}^{S0}$ $N_{S3_1}^{S1}$ $N_{S3_1}^{S0}$ $N_{S3_2}^{S1}$ and $N_{S3_2}^{S0}$ 
increases with the rise in $NC_{prob}$ according to the $P(\exists NC)$ analysis earlier. Also, as $n_{veh}$ rises, this denotes that there will be more competing CAVs to HV and thus the percentages of the above-mentioned transitions increase even further. 

In addition, it can also be observed that for $n_{veh}=$ {8, 16}, the heat maps display a higher percentages for 
$N_{S3_1}^{S1}$
and
$N_{S3_2}^{S1}$
as compared to 
$N_{S3_1}^{S0}$
and
$N_{S3_2}^{S0}$
respectively. As $n_{veh}$ increases, so does the percentage of 
$N_{S1}^{SW}$
since there is a higher probability that HV would encounter CAVs competing for AROW when it arrives at the SC-I. This in turn explains the increase in the percentages of 
$N_{SW}^{S3_1}$
and
$N_{SW}^{S3_2}$
where HV goes from $SW$ directly to $S3_1$ or $S3_2$ towards the end of the successful preceding AROW. Recalling the logic behind $AROW_{thresh}$, this means that as the HV arrives at $S3_1$ or $S3_2$, its $AROW_{thresh}$ is still 0, and thus if there is non-compliance and (\ref{eq_arowthresh}) holds true, HV would restart its AROW and use 
$N_{S3_1}^{S1}$
and
$N_{S3_2}^{S1}$
to transition to $S1$ rather than transitioning to $S0$ and disable its AROW. This behavior can be observed for both non-zero $NC_{prob}$ values. 
For the same reason, the percentages of 
$N_{S2_2}^{S0}$
and
$N_{S2_3}^{S0}$
are higher than that of 
$N_{S2_2}^{S1}$
and 
$N_{S2_3}^{S1}$
because as HV reaches $S2_2$ or $S2_3$, it is probable that due to the non-compliance in the previous AROW round, the $AROW_{count}$ is close to reaching $AROW_{thresh}$, and thus a non-compliance here would mean that HV would disable its AROW.
These behaviors are not viewable in the case of $n_{veh}$ = 4 simply because there are not enough CAVs in the scenario to force HV to transition to $SW$ upon arriving at the SC-I. 

Finally, Figure \ref{fig:arow_nc_statetransitions} also shows that for any $n_{veh}$, the percentage of 
$N_{S3_1}^{S4_1}$
and 
$N_{S3_2}^{S4_2}$
reduces with the increase of $NC_{prob}$ and that is a direct consequence of $NC_{prob}$ as it reduces the probability of a successful AROW completion.

\paragraph{Significance of the convergence of $P_E(\exists NC)$ with $P_T(\exists NC)$}

As shown in Figure \ref{fig:atleast1NC_theo_exp}, the convergence of $P_E(\exists NC)$ with $P_T(\exists NC)$
under varying levels of non-compliance at different stages of AROW and with different number of competing CAVs
can be very useful in the context of AROW.
Firstly, it allows further testing of different AROW scenarios in the future using the theoretical calculations as a proxy without explicitly running the simulations and obtaining experimental data. This can potentially accelerate the development and scalability of AROW as an application for CAVs on commercial roads.
Secondly, the theoretical analysis obtained from any given test scenario of AROW can be utilized as a reference to detect any anomalies whether it is due to the implementation of that scenario or any specific CAV agent and its behavior.
\\
\subsubsection{AROW Impact based on Evaluation Metrics}
The impact of AROW is evaluated on the basis of ambiguity mitigation and fairness. 
The main contribution of AROW is to reduce the ambiguity that drivers may face upon arriving at the SC-I at close time intervals to each other. This in turn can allow fairness in terms of SC-I crossing priority assignments. In this regard, the two metrics are mentioned below:
\begin{itemize}
    \item For ambiguity reduction, we calculate the frequency of ambiguous instances ($n_{amb}$) at SC-I where vehicles arrive at close intervals of each other. $n_{amb}$ is characterized by the number of false starts by vehicles in situations where the right-of-way is not clearly evident.
    \item For fairness-based metric, the vehicle clearance time ($T_{clearance}$) is calculated for a set of leading competing vehicles 
    at an intersection that arrive within $T1$ of each other. This is the time difference between the first vehicle in 
    the set 
    entering the junction and the last vehicle in 
    the set
    exiting the junction. SC-I junction can be described as the region within the SC-I past the stop sign. $T_{clearance}$ can be described as:
    \begin{equation}
        T_{clearance} = \max(D_{T_{exit}}) - \min(D_{T_{arrive}})
    \end{equation}
    where 
    $D_{T_{arrive}}$
    and 
    $D_{T_{exit}}$
    are the sets for the junction arrival and departure timestamp of the set of vehicles, respectively.
\end{itemize}

For $n_{amb}$, AROW is compared to
our implementation of the algorithm in \cite{hassan2014fully}, which we refer to as Delay-based Cooperative Intersection Management (DCIM), and  
the allway stop intersection management method from SUMO. 
The allway method is similar to the real-life human behavior at an SC-I. The default rules of allway stop within SUMO are according to the general right-of-way rules \cite{mutcd}. 
To avoid any confusion, it should be noted that only those vehicles utilizing AROW 
and DCIM
are denoted as CAVs.
Table \ref{table:metric_allwayconflict} provides a tabular comparison of $n_{amb}$ instances of AROW, DCIM, and allway.
This table contains percentages of competing vehicles entering the SC-I together from different lanes. There are three different instances of competing vehicles simultaneously entering the SC-I, i.e. 1, 2, or 3. All these instances are represented as a separate column for each of the AROW, DCIM, and allway scenarios. In the case of 1 competing vehicle, this includes all instances where only a single vehicle entered the SC-I in foucs, thus in such cases, there are no $n_{amb}$ instances. On the other hand, the columns of 2 and 3 denote the cases of 2 and 3 competing vehicles entering simultaneously within an SC-I as a result of ambiguity, thus, these instances are denoted as $n_{amb}$ in Table \ref{table:metric_allwayconflict}. For each $n_{veh}$, there is one scenario for each of DCIM, and allway, whereas there are three scenarios for AROW for each of the non-compliance values, i.e. $NC_{prob} = 0, 0.1,$ and 0.25.

\begin{table}[b]
  \centering
  \caption{\small{Tabular comparison of SC-I scenarios with false-starts due to ambiguity ($n_{amb}$) using AROW, allway, and DCIM. Each method contains instances with 1, 2, or 3 competing vehicles and instances of 2 and 3 CAVs constitute $n_{amb}$ cases.}}
  \resizebox{0.48\textwidth}{!}{
  \begin{tabular}{|c|c|c|c|c|c|c|c|c|c|c|}
    \hline
    \multirow{4}{*}{\boldmath{$n_{veh}$}} &
     \multirow{3}{*}{} &
     \multicolumn{9}{c|}{\normalsize{Competing Vehicles inside SC-I (\%)}}\\\cline{3-11} &
     \multicolumn{1}{c|}{} &
     \multicolumn{3}{c|}{AROW} &
     \multicolumn{3}{c|}{DCIM} & 
     \multicolumn{3}{c|}{Allway}\\\cline{3-11} &
     \multicolumn{1}{c|}{} &
     \multicolumn{1}{c|}{} &
     \multicolumn{2}{c|}{\boldmath{$n_{amb}$}} &
     \multicolumn{1}{c|}{} &
     \multicolumn{2}{c|}{\boldmath{$n_{amb}$}} & 
     \multicolumn{1}{c|}{} &
     \multicolumn{2}{c|}{\boldmath{$n_{amb}$}}\\\cline{2-11} & 
     \textbf{$NC_{prob}$} & \textbf{1} & \textbf{2} & \textbf{3} & \textbf{1} & \textbf{2} & \textbf{3} & \textbf{1} & \textbf{2} & \textbf{3}\\
     \hline
    \multirow{3}{*}{\textbf{4}} & \textbf{0} & 100 & 0  & 0 & & & & & &\\\cline{2-5}
    & \textbf{0.1} & 100 & 0  & 0 & 97.3 & 2.7  & 0 & 96 & 3.8  & 0.2\\\cline{2-5}
    & \textbf{0.25} & 99.8 & 0.2 & 0 & & & & & &\\
   \hline  
    \multirow{3}{*}{\textbf{8}}  & \textbf{0} & 100 & 0  & 0 & & & & & &\\\cline{2-5}
    & \textbf{0.1} & 99.4 & 0.5  & 0.1 & 91.6 & 8.3  & 0.1 & 84.5 & 14.8  & 0.7\\\cline{2-5}
    & \textbf{0.25} & 96.9 & 2.5 & 0.6 & & & & & &\\
   \hline
       \multirow{3}{*}{\textbf{16}}  & \textbf{0} & 100 & 0  & 0 & & & & & &\\\cline{2-5}
    & \textbf{0.1} & 97.8 & 0.9  & 1.3 & 86.1 & 13.5  & 0.4 & 82.2 & 16.9  & 0.9\\\cline{2-5}
    & \textbf{0.25} & 89.9 & 7.2 & 2.9 & & & & & &\\
   \hline
  \end{tabular}
  }
  \label{table:metric_allwayconflict}
\end{table}

On a high-level, it can be observed that AROW has the least percentage of false-start instances due to ambiguity, i.e. $n_{amb}$, followed by DCIM and then the default allway, which performs the worst.
In the case of DCIM, since its main focus is on time delay reduction and not ambiguity mitigation as mentioned in Section \ref{Section1}, there are limitations in its logic in assigning priorities as described in the $T_{clearance}$ comparison that can induce ambiguity. On such related limitation is that DCIM can assign similar priorities to multiple vehicles from different lanes having the same traffic load and SC-I arrival timestamp. This limitation is the reason behind the increase in $n_{amb}$ instances using DCIM as compared to AROW. However, DCIM still outperforms the default allway scenarios since it is still a V2X based algorithm where cooperation among CAVs reduces numerous $n_{amb}$ instance which otherwise occur using the non-cooperative default allway that have the most $n_{amb}$ instances.

\begin{figure*}
\centering
\includegraphics[trim=100 0 100 0,width=0.75\textwidth]
{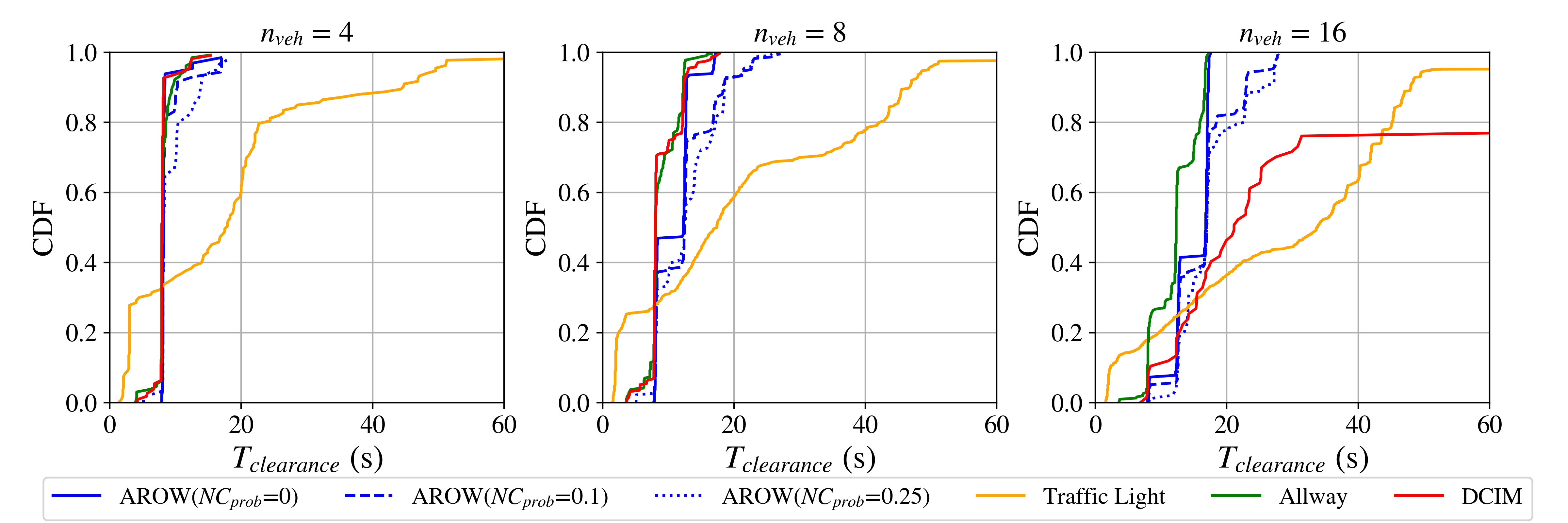}
\caption{\small{CDF comparison of vehicle clearance time ($T_{clearance}$) between AROW, traffic light, and allway.}}
\label{fig:metric_CDFclearance}
\end{figure*}

More specifically, Table \ref{table:metric_allwayconflict} shows that for any $n_{veh}$, in the case of AROW with $NC_{prob} = 0$, instances with 1 competing CAV within SC-I at any time throughout the simulations constitute for 100\% of the scenarios, thus no $n_{amb}$. However, this is not the case with non-zero $NC_{prob} = 0.1, 0.25$ where we observe some $n_{amb}$ instances.
Non-zero $NC_{prob}$ means that there can be cases of repetitive non-compliance which can satisfy (\ref{eq_arowthresh}) and disable AROW. Once AROW is disabled, the remaining compliant CAVs can follow any default heuristic to manually pass the SC-I. For this study, the default heuristic is set to FIFO which is usual for real-life non-cooperative SC-I scenarios. However, such scenarios with CAVs manually passing the SC-I can cause ambiguity and thus, $n_{amb}$ instances.
Given this, as $NC_{prob}$ rises, we should experience higher cases of disabled AROW and higher $n_{amb}$ instances with 2 or 3 competing CAVs entering as a false start due to ambiguity, which is precisely what we observe in Table \ref{table:metric_allwayconflict}. 

However, despite this, the percentage of $n_{amb}$ instances within AROW is still far lower as compare to allway and DCIM.
In the case of $n_{veh} = 4$, there is 0.2\% instances of 2 competing CAVs when $NC_{prob} = 0.25$, while for the same $n_{veh} = 4$, DCIM has 2.7\% of $n_{amb}$ cases which further rises to 4\% for allway with 3.8\% and 0.2\% of cases with 2 and 3 competing vehicles, respectively.
Similarly, in the case of $n_{veh} = 8$, AROW has a total $n_{amb}$ instances of 0.6\% in the case of $NC_{prob} = 0.1$ which rises to only 3.1\% with $NC_{prob} = 0.25$. This is far lesser when compared to DCIM for $n_{veh} = 8$ which contains a total of 8.4\% of $n_{amb}$ scenarios. And as expected, allway performs the poorest with a total of 15.5\% of $n_{amb}$. Same trend can be observed in the case of $n_{veh} = 16$ where AROW encounters the least amount of $n_{amb}$ even when it is faced with a high $NC_{prob}$, followed by DCIM which performs better than the default non-cooperative allway.


In addition to ambiguity reduction, AROW also promotes fairness among competing vehicles at an SC-I without necessarily causing additional delays. Figure \ref{fig:metric_CDFclearance} compares the $T_{clearance}$ CDF of AROW to 
DCIM, 
along with allway and traffic signal methods from SUMO. 
The results from traffic light based intersection have been included to serve as a baseline. 
For $n_{veh} = 4, 8$, the $T_{clearance}$ of AROW with $NC_{prob} = 0$, DCIM, and allway are similar to each other and they significantly outperform traffic light-based intersection. With $NC_{prob} = 0.1, 0.25$, there appears to be a slight degradation in $T_{clearance}$ for AROW, which is expected since compliant CAVs re-run AROW in the face of non-compliance until (\ref{eq_arowthresh}) is satisfied. And the tradeoff for this as explained above is the reduction in $n_{amb}$ instances for AROW as compared to other methods.
In the case of high density traffic ($n_{veh} = 16$), the $T_{clearance}$ of AROW with $NC_{prob}=0$ and allway are still similar and as expected, the $T_{clearance}$ of AROW slightly degrades with $NC_{prob} = 0.1, 0.25$. However, there is a significant deterioration in the $T_{clearance}$ of DCIM, which is due to its limitations as explained below.

Firstly, 
DCIM aims to replace SC-I by assigning time-delaying instructions to CAVs so that they can varyingly decelerate and arrive at the intersection at different times, thereby crossing the intersection without stopping to let other CAVs pass. Although this logic can work in low and medium density traffic scenarios where CAVs have enough distance headway to decelerate and arrive at the SC-I when it is not occupied, however, the same is not the case in high density traffic
with limited or no headway, 
where CAVs will most likely encounter higher priority CAVs at SC-I and thus, would have to stop and wait. This causes the CAVs in DCIM to possess a higher time delay which partly contributes to the deterioration of $T_{clearance}$ for $n_{veh}=$ 16 as visible in Figure \ref{fig:metric_CDFclearance}. 
Secondly, to enable smooth traffic flow and reduce time delay, DCIM also contains the logic of assigning a higher 
intersection crossing 
priority to CAVs that belong to lane with a higher CAV load. This logic is also limited since for high density and uneven traffic as for $n_{veh}=$ 16, CAVs belonging to lesser priority lanes can potentially wait for a long time until their lane can get priority to pass, which causes the $T_{clearance}$ to rise.
Thirdly, in the case of conflict where multiple CAVs arrive at SC-I at the same time despite the scheduling, a scenario that can occur quite often for $n_{veh}=$ 16, DCIM follows the same principle of giving priority to CAVs in the higher-priority lanes and it schedules the lesser priority CAVs at the earliest possible empty slot. This is also a major limitation of DCIM since in the case of high density traffic where there are limited or no empty slots at SC-I, a lesser priority CAV can potentially wait at the SC-I for a long time and cause a high $T_{clearance}$.  
Thus, in terms of fairness which is the basis of the metric $T_{clearance}$ in this study, DCIM is not suitable while AROW can easily handle high density traffic and remain fair to CAVs based on their arrival at the SC-I, while mitigating ambiguity.

\section{Conclusion}
This study proposes a 
V2X-based distributed
AROW algorithm
for CIM
to reduce priority-related ambiguity at SC-Is. The algorithm utilizes cooperative DMS where CAVs use V2X to communicate with each other
and negotiate SC-I crossing priorities in the form of DIMs. 
AROW is scalable for high density traffic and is capable of handling varying levels of driver non-compliance towards assigned priorities.
We provide details for all stages within AROW and also implement it in terms of a timed automaton.
Later, we run several simulation scenarios involving various levels of CAVs and non-compliance probabilities to prove the completeness and robustness of AROW. 
The paper iterates the accuracy of the proposed algorithm by showing close similarities between the theoretical calculations and experimental results.
Finally, AROW is evaluated on the basis of ambiguity mitigation and intersection clearance time as compared to 
a related work (DCIM), 
traffic light intersection, and allway based SC-I from SUMO. 
In terms of ambiguity mitigation,
AROW is 
observed to outperform all the other methods for all CAVs and non-compliance scenarios.
In terms of clearance time, AROW outperforms DCIM and traffic lights for all scenarios while it performs similar to allway with no non-compliance and a slight comparative degradation in non-compliance scenarios due to compliant CAVs within AROW restarting the process in the face of non-compliance by any of the participating CAVs.



\section{Future Works}
We plan to test AROW under real-life conditions by embedding the DMS and the underlying AROW on to the test hardware. 
Secondly, AROW can be stress-tested under varying levels of congested channels and communication losses to further improve the robustness of the algorithm. 
Current study assumes full connectivity, which can be relaxed to show the robustness of AROW in the presence of connected and non-connected vehicles 
along with CAVs that have disabled AROW. 
Thirdly, we can also incorporate emergency CAVs in the simulations and evaluate the performance with respect to the relevant metrics since the core logic of AROW is capable of handling CAVs with different priorities. 
Fourthly, AROW performance against non-compliance can also be potentially improved by learning the non-compliance behavior of different CAVs and predicting non-compliance in any given SC-I scenario.
Finally, the timeouts $T1$, $T2$, and $T3$ can be optimized by conducting analysis on communication latencies using different V2X technologies. 

\vspace{11pt}


\bibliographystyle{IEEEtran}
\bibliography{main.bib}{}

\begin{thebibliography}{10}
\providecommand{\url}[1]{#1}
\csname url@samestyle\endcsname
\providecommand{\newblock}{\relax}
\providecommand{\bibinfo}[2]{#2}
\providecommand{\BIBentrySTDinterwordspacing}{\spaceskip=0pt\relax}
\providecommand{\BIBentryALTinterwordstretchfactor}{4}
\providecommand{\BIBentryALTinterwordspacing}{\spaceskip=\fontdimen2\font plus
\BIBentryALTinterwordstretchfactor\fontdimen3\font minus \fontdimen4\font\relax}
\providecommand{\BIBforeignlanguage}[2]{{%
\expandafter\ifx\csname l@#1\endcsname\relax
\typeout{** WARNING: IEEEtran.bst: No hyphenation pattern has been}%
\typeout{** loaded for the language `#1'. Using the pattern for}%
\typeout{** the default language instead.}%
\else
\language=\csname l@#1\endcsname
\fi
#2}}
\providecommand{\BIBdecl}{\relax}
\BIBdecl

\bibitem{rafter2017traffic}
C.~B. Rafter, B.~Anvari, and S.~Box, ``Traffic responsive intersection control algorithm using gps data,'' in \emph{2017 IEEE 20th International Conference on Intelligent Transportation Systems (ITSC)}.\hskip 1em plus 0.5em minus 0.4em\relax IEEE, 2017, pp. 1--6.

\bibitem{rahmati2017towards}
Y.~Rahmati and A.~Talebpour, ``Towards a collaborative connected, automated driving environment: A game theory based decision framework for unprotected left turn maneuvers,'' in \emph{2017 IEEE Intelligent Vehicles Symposium (IV)}.\hskip 1em plus 0.5em minus 0.4em\relax IEEE, 2017, pp. 1316--1321.

\bibitem{blincoe2015economic}
L.~Blincoe, T.~R. Miller, E.~Zaloshnja, and B.~A. Lawrence, ``The economic and societal impact of motor vehicle crashes, 2010 (revised),'' Tech. Rep., 2015.

\bibitem{retting2003analysis}
R.~A. Retting, H.~B. Weinstein, and M.~G. Solomon, ``Analysis of motor-vehicle crashes at stop signs in four us cities,'' \emph{Journal of Safety Research}, vol.~34, no.~5, pp. 485--489, 2003.

\bibitem{shi2016intelligent}
M.-k. Shi, H.~Jiang, and S.-h. Li, ``An intelligent traffic-flow-based real-time vehicles scheduling algorithm at intersection,'' in \emph{2016 14th International Conference on Control, Automation, Robotics and Vision (ICARCV)}.\hskip 1em plus 0.5em minus 0.4em\relax IEEE, 2016, pp. 1--5.

\bibitem{choi2010crash}
E.-H. Choi, ``Crash factors in intersection-related crashes: An on-scene perspective,'' 2010.

\bibitem{azimi2014stip}
R.~Azimi, G.~Bhatia, R.~R. Rajkumar, and P.~Mudalige, ``Stip: Spatio-temporal intersection protocols for autonomous vehicles,'' in \emph{2014 ACM/IEEE international conference on cyber-physical systems (ICCPS)}.\hskip 1em plus 0.5em minus 0.4em\relax IEEE, 2014, pp. 1--12.

\bibitem{singh2015critical}
S.~Singh, ``Critical reasons for crashes investigated in the national motor vehicle crash causation survey,'' Tech. Rep., 2015.

\bibitem{chen2015cooperative}
L.~Chen and C.~Englund, ``Cooperative intersection management: A survey,'' \emph{IEEE transactions on intelligent transportation systems}, vol.~17, no.~2, pp. 570--586, 2015.

\bibitem{namazi2019intelligent}
E.~Namazi, J.~Li, and C.~Lu, ``Intelligent intersection management systems considering autonomous vehicles: A systematic literature review,'' \emph{IEEE Access}, vol.~7, pp. 91\,946--91\,965, 2019.

\bibitem{zhong2020autonomous}
Z.~Zhong, M.~Nejad, and E.~E. Lee, ``Autonomous and semiautonomous intersection management: A survey,'' \emph{IEEE Intelligent Transportation Systems Magazine}, vol.~13, no.~2, pp. 53--70, 2020.

\bibitem{gholamhosseinian2022comprehensive}
A.~Gholamhosseinian and J.~Seitz, ``A comprehensive survey on cooperative intersection management for heterogeneous connected vehicles,'' \emph{IEEE Access}, vol.~10, pp. 7937--7972, 2022.

\bibitem{jkenney:dsrcmain}
J.~B. Kenney, ``Dedicated short-range communications (dsrc) standards in the united states,'' \emph{Proceedings of the IEEE}, vol.~99, no.~7, pp. 1162--1182, July 2011.

\bibitem{3gpp:3gppmain}
3GPP, ``Evolved universal terrestrial radio access (e-utra); physical layer procedures (v14.7.0),'' vol. 3GPP. TS 36.213, June 2018.

\bibitem{shah2019real}
G.~Shah, R.~Valiente, N.~Gupta, S.~O. Gani, B.~Toghi, Y.~P. Fallah, and S.~D. Gupta, ``Real-time hardware-in-the-loop emulation framework for dsrc-based connected vehicle applications,'' in \emph{2019 IEEE 2nd Connected and Automated Vehicles Symposium (CAVS)}.\hskip 1em plus 0.5em minus 0.4em\relax IEEE, 2019, pp. 1--6.

\bibitem{shah2020rve}
G.~Shah, M.~Saifuddin, Y.~P. Fallah, and S.~D. Gupta, ``Rve-cv2x: A scalable emulation framework for real-time evaluation of cv2x-based connected vehicle applications,'' in \emph{2020 IEEE Vehicular Networking Conference (VNC)}.\hskip 1em plus 0.5em minus 0.4em\relax IEEE, 2020, pp. 1--8.

\bibitem{schmocker2008multi}
J.-D. Schm{\"o}cker, S.~Ahuja, and M.~G. Bell, ``Multi-objective signal control of urban junctions--framework and a london case study,'' \emph{Transportation Research Part C: Emerging Technologies}, vol.~16, no.~4, pp. 454--470, 2008.

\bibitem{devoe2008distributed}
D.~DeVoe and R.~W. Wall, ``A distributed smart signal architecture for traffic signal controls,'' in \emph{2008 IEEE International Symposium on Industrial Electronics}.\hskip 1em plus 0.5em minus 0.4em\relax IEEE, 2008, pp. 2060--2065.

\bibitem{chang2013study}
H.-J. Chang and G.-T. Park, ``A study on traffic signal control at signalized intersections in vehicular ad hoc networks,'' \emph{Ad Hoc Networks}, vol.~11, no.~7, pp. 2115--2124, 2013.

\bibitem{buinevich2021v2x}
M.~Buinevich, A.~Spirkina, V.~Elagin, S.~Tarakanov, and A.~Vladyko, ``V2x-based intersection priority management,'' in \emph{2021 Systems of Signals Generating and Processing in the Field of on Board Communications}.\hskip 1em plus 0.5em minus 0.4em\relax IEEE, 2021, pp. 1--7.

\bibitem{gutesa2021development}
S.~Gutesa, J.~Lee, and D.~Besenski, ``Development and evaluation of cooperative intersection management algorithm under connected and automated vehicles environment,'' \emph{Transportation research record}, vol. 2675, no.~7, pp. 94--104, 2021.

\bibitem{parks2022intersection}
A.~Parks-Young and G.~Sharon, ``Intersection management protocol for mixed autonomous and human-operated vehicles,'' \emph{IEEE Transactions on Intelligent Transportation Systems}, vol.~23, no.~10, pp. 18\,315--18\,325, 2022.

\bibitem{dresner2004multiagent}
K.~Dresner and P.~Stone, ``Multiagent traffic management: A reservation-based intersection control mechanism,'' in \emph{Autonomous Agents and Multiagent Systems, International Joint Conference on}, vol.~3.\hskip 1em plus 0.5em minus 0.4em\relax Citeseer, 2004, pp. 530--537.

\bibitem{kowshik2011provable}
H.~Kowshik, D.~Caveney, and P.~Kumar, ``Provable systemwide safety in intelligent intersections,'' \emph{IEEE transactions on vehicular technology}, vol.~60, no.~3, pp. 804--818, 2011.

\bibitem{perronnet2012cooperative}
F.~Perronnet, A.~Abbas-Turki, J.~Buisson, A.~El~Moudni, R.~Zeo, and M.~Ahmane, ``Cooperative intersection management: Real implementation and feasibility study of a sequence based protocol for urban applications,'' in \emph{2012 15th international IEEE conference on intelligent transportation systems}.\hskip 1em plus 0.5em minus 0.4em\relax IEEE, 2012, pp. 42--47.

\bibitem{wu2012cooperative}
J.~Wu, A.~Abbas-Turki, and A.~El~Moudni, ``Cooperative driving: an ant colony system for autonomous intersection management,'' \emph{Applied Intelligence}, vol.~37, pp. 207--222, 2012.

\bibitem{li2020intersection}
Y.~Li and Q.~Liu, ``Intersection management for autonomous vehicles with vehicle-to-infrastructure communication,'' \emph{PLoS one}, vol.~15, no.~7, p. e0235644, 2020.

\bibitem{gregoire2016hybrid}
J.~Gregoire and E.~Frazzoli, ``Hybrid centralized/distributed autonomous intersection control: Using a job scheduler as a planner and inheriting its efficiency guarantees,'' in \emph{2016 IEEE 55th Conference on Decision and Control (CDC)}.\hskip 1em plus 0.5em minus 0.4em\relax IEEE, 2016, pp. 2549--2554.

\bibitem{vanmiddlesworth2008replacing}
M.~VanMiddlesworth, K.~Dresner, and P.~Stone, ``Replacing the stop sign: Unmanaged intersection control for autonomous vehicles,'' in \emph{Proceedings of the 7th international joint conference on Autonomous agents and multiagent systems-Volume 3}, 2008, pp. 1413--1416.

\bibitem{li2006cooperative}
L.~Li and F.-Y. Wang, ``Cooperative driving at blind crossings using intervehicle communication,'' \emph{IEEE Transactions on Vehicular technology}, vol.~55, no.~6, pp. 1712--1724, 2006.

\bibitem{katriniok2019nonlinear}
A.~Katriniok, P.~Sopasakis, M.~Schuurmans, and P.~Patrinos, ``Nonlinear model predictive control for distributed motion planning in road intersections using panoc,'' in \emph{2019 IEEE 58th Conference on Decision and Control (CDC)}.\hskip 1em plus 0.5em minus 0.4em\relax IEEE, 2019, pp. 5272--5278.

\bibitem{hassan2014fully}
A.~A. Hassan and H.~A. Rakha, ``A fully-distributed heuristic algorithm for control of autonomous vehicle movements at isolated intersections,'' \emph{International Journal of Transportation Science and Technology}, vol.~3, no.~4, pp. 297--309, 2014.

\bibitem{mutcd}
\BIBentryALTinterwordspacing
{U.S. Department of Transportation, Federal Highway Administration}, ``Manual on uniform traffic control devices (mutcd),'' Online, 2003, accessed on March 26, 2023. [Online]. Available: \url{https://mutcd.fhwa.dot.gov/htm/2003r1/part2/part2b1.htm#section2B04}
\BIBentrySTDinterwordspacing

\bibitem{shah2022enabling}
G.~Shah, S.~Shahram, Y.~Fallah, D.~Tian, and E.~Moradi-Pari, ``Enabling a cooperative driver messenger system for lane change assistance application,'' in \emph{2022 IEEE 25th International Conference on Intelligent Transportation Systems (ITSC)}.\hskip 1em plus 0.5em minus 0.4em\relax IEEE, 2022, pp. 4320--4327.

\bibitem{tian2023systems}
D.~Tian, E.~M. Pari, G.~Shah, S.~Shahram, and Y.~P. Fallah, ``Systems and methods for a point-to-point driver messenger for advanced cooperative situational awareness,'' Sep.~21 2023, uS Patent App. 18/184,446.

\bibitem{rajab2017driver}
S.~Rajab, S.~Bai, S.~Saigusa, J.~Keller, A.~Pradhan, S.~Bao, and J.~Sullivan, ``Driver to driver (d2d) personalized messaging based on connected vehicles: Concept evaluation via simulation,'' 2017.

\bibitem{alur1994theory}
R.~Alur and D.~L. Dill, ``A theory of timed automata,'' \emph{Theoretical computer science}, vol. 126, no.~2, pp. 183--235, 1994.

\bibitem{alur1999timed}
R.~Alur, ``Timed automata,'' in \emph{International Conference on Computer Aided Verification}.\hskip 1em plus 0.5em minus 0.4em\relax Springer, 1999, pp. 8--22.

\bibitem{krauss1997metastable}
S.~Krau{\ss}, P.~Wagner, and C.~Gawron, ``Metastable states in a microscopic model of traffic flow,'' \emph{Physical Review E}, vol.~55, no.~5, p. 5597, 1997.

\bibitem{SUMO2018}
\BIBentryALTinterwordspacing
P.~A. Lopez, M.~Behrisch, L.~Bieker-Walz, J.~Erdmann, Y.-P. Fl{\"o}tter{\"o}d, R.~Hilbrich, L.~L{\"u}cken, J.~Rummel, P.~Wagner, and E.~Wie{\ss}ner, ``Microscopic traffic simulation using sumo,'' in \emph{The 21st IEEE International Conference on Intelligent Transportation Systems}.\hskip 1em plus 0.5em minus 0.4em\relax IEEE, 2018. [Online]. Available: \url{https://elib.dlr.de/124092/}
\BIBentrySTDinterwordspacing

\bibitem{wegener2008traci}
A.~Wegener, M.~Pi{\'o}rkowski, M.~Raya, H.~Hellbr{\"u}ck, S.~Fischer, and J.-P. Hubaux, ``Traci: an interface for coupling road traffic and network simulators,'' in \emph{Proceedings of the 11th communications and networking simulation symposium}, 2008, pp. 155--163.

\bibitem{saej3161}
S.~International, ``Lte vehicle-to-everything (lte-v2x) deployment profiles and radio parameters for single radio channel multi-service coexistence,'' Society of Automotive Engineers, Standard J3161, 04 2022.

\bibitem{toghi2018multiple}
B.~Toghi, M.~Saifuddin, H.~N. Mahjoub, M.~O. Mughal, Y.~P. Fallah, J.~Rao, and S.~Das, ``Multiple access in cellular v2x: Performance analysis in highly congested vehicular networks,'' in \emph{2018 IEEE Vehicular Networking Conference (VNC)}.\hskip 1em plus 0.5em minus 0.4em\relax IEEE, 2018, pp. 1--8.

\end{thebibliography}


\vskip -2.5\baselineskip plus -1fil
\begin{IEEEbiography}[{\includegraphics[width=1in,height=1.25in,clip,keepaspectratio]{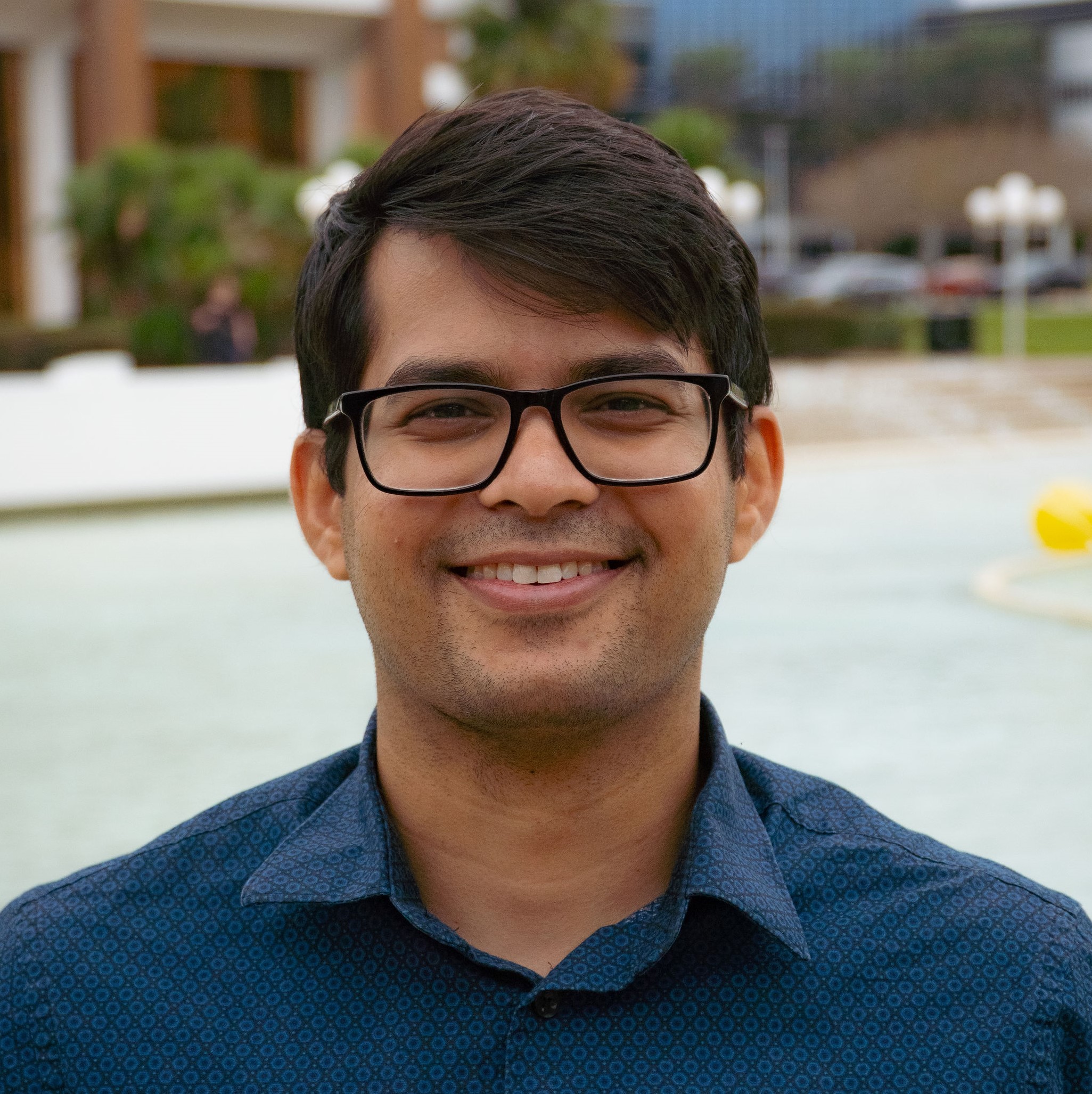}}]{Ghayoor Shah}
is a Ph.D. candidate at the University of Central Florida. He received the B.Sc. degree in Computer Engineering from University of Illinois at Urbana-Champaign in 2018. He has previously worked as a mobility engineering intern at Phantom Auto and as a research intern at Ford Motor Company. 
His research interests include Connected and Autonomous Vehicles (CAVs), scalability analysis of V2X, and applications of artificial intelligence to cooperative driving. 
\end{IEEEbiography}
\vskip -2.5\baselineskip plus -1fil
\begin{IEEEbiography}[{\includegraphics[width=1in,height=1.25in,clip,keepaspectratio]{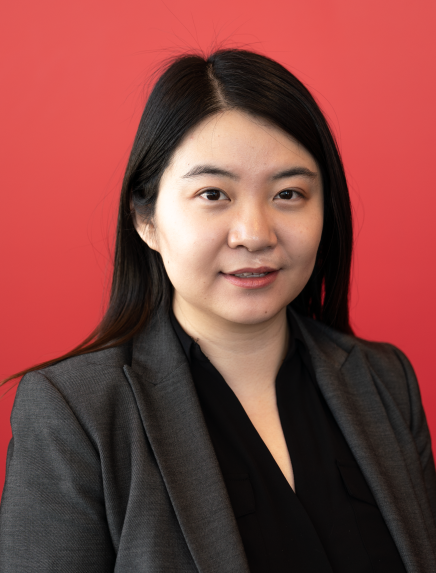}}]{Danyang Tian} earned her Ph.D. degree in electrical engineering from the University of California, Riverside, in 2018. She is currently a scientist at Honda Research Institute USA, Inc. Her research interests include connected and automated vehicles, vehicular communication systems, and driver behavior modeling, application design and development of connected and automated vehicles, vehicle-to-everything (V2X) communication technologies. She serves as chair in the SAE International V2X Vehicular Applications Technical Committee.
\end{IEEEbiography}
\vskip -2.5\baselineskip plus -1fil
\begin{IEEEbiography}[{\includegraphics[width=1in,height=1.25in,clip,keepaspectratio]{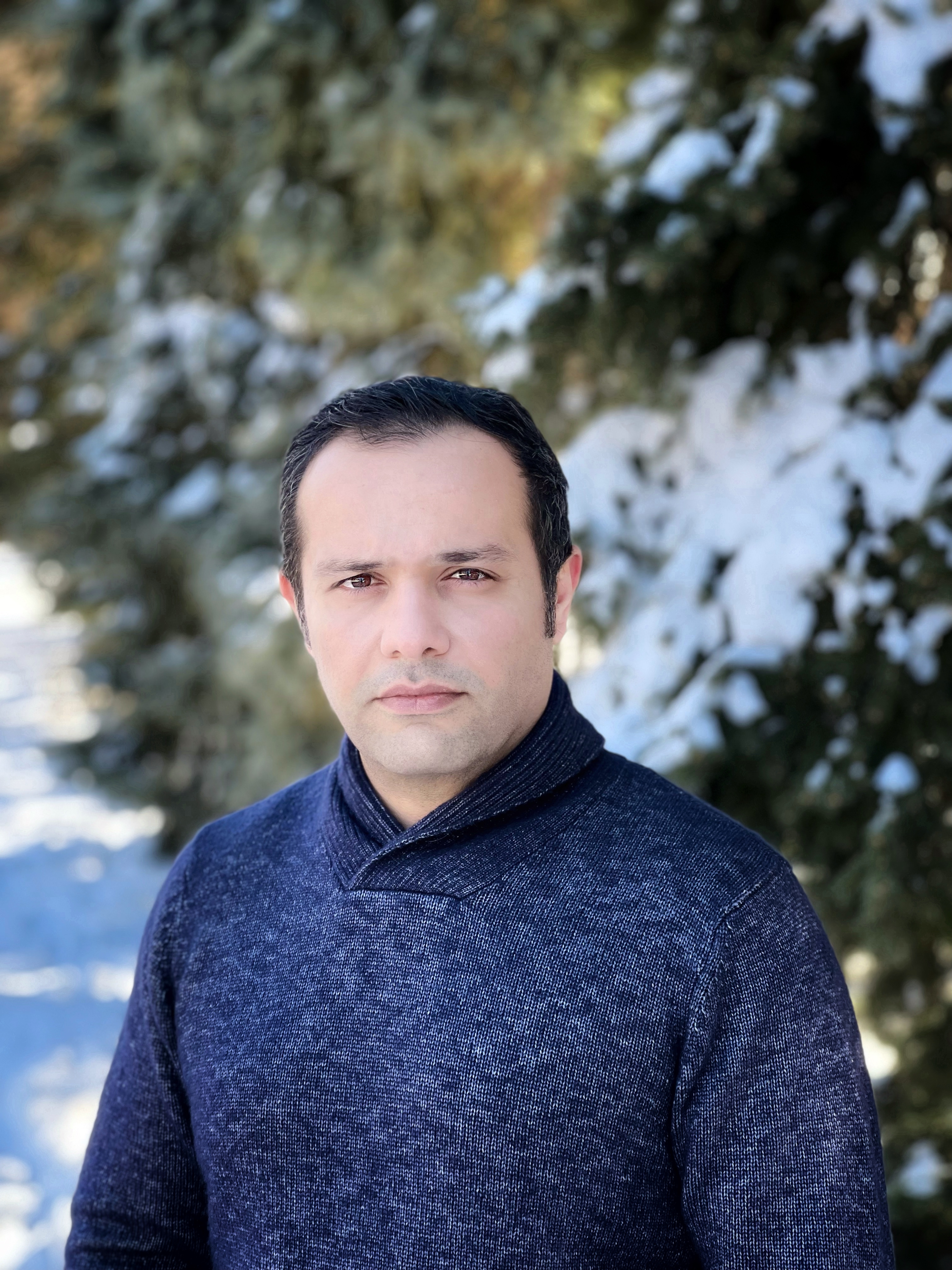}}]{Ehsan Moradi-Pari} is a lead principal scientist and communication/data research group lead at Honda Research Institute (HRI-US). His current research is focused on vehicle-to-vehicle (V2V) and vehicle-to-infrastructure (V2I) communications based on short-range communication and cellular technologies. He serves as Honda lead and representative for V2V and V2I precompetitive research such as Vehicle-to-Vehicle Systems Engineering and Vehicle Integration Research for Deployment, Cooperative Adaptive Cruise Control, and 5G Automotive Association (5GAA). He has spearheaded Honda’s efforts and preparations to accelerate connected automated vehicle deployment in the United States and the state of Ohio where he led the team to design and deploy the first-of-its-kind smart intersection concept in Marysville, Ohio. Dr. Moradi-Pari has several publications in internationally refereed journals, conference proceedings, and book chapters in the area of designing communication protocols and applications for connected and automated vehicles and modeling/analysis of intelligent transportation systems
\end{IEEEbiography}
\vskip -2.5\baselineskip plus -1fil
\begin{IEEEbiography}[{\includegraphics[width=1in,height=1.25in,clip,keepaspectratio]{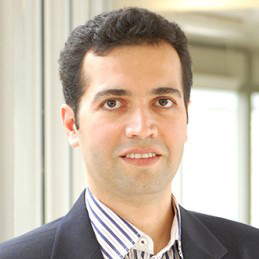}}]{Yaser P. Fallah} is an Associate Professor in the ECE Department at the University of Central Florida. He received the Ph.D. degree from the University of British Columbia, Vancouver, BC, Canada, in 2007. From 2008 to 2011, he was a Research Scientist with the Institute of Transportation Studies,
University of California Berkeley, Berkeley, CA, USA. His research, sponsored by industry, USDoT, and NSF, is focused on intelligent transportation systems and automated and networked vehicle safety systems.
\end{IEEEbiography}

\end{document}